\documentclass[prl,reprint,twocolumn,showpacs,superscriptaddress]{revtex4-1}
\usepackage{amsmath,amssymb,bm,mathrsfs,graphicx, braket,amsthm,enumerate,times}
\usepackage[colorlinks=true,citecolor=blue,linkcolor=blue]{hyperref}
\usepackage{longtable}
\usepackage{multirow}

\renewcommand{\vec}[1]{\bm{#1}}

\begin{document}
\title{Universal Relation among Many-Body Chern Number, Rotation Symmetry, and Filling}
\begin{abstract}
Understanding the interplay between the topological nature and the symmetry property of interacting systems has been a central matter of condensed matter physics in recent years.
In this Letter, we establish nonperturbative constraints on the quantized Hall conductance of many-body systems with arbitrary interactions.  Our results allow one to readily determine the many-body Chern number modulo a certain integer without performing any integrations, solely based on the rotation eigenvalues and the average particle density of the many-body ground state.
\end{abstract}

\author{Akishi Matsugatani} 
\affiliation{Department of Applied Physics, University of Tokyo, Tokyo 113-8656, Japan}

\author{Yuri Ishiguro} 
\affiliation{Department of Applied Physics, University of Tokyo, Tokyo 113-8656, Japan}

\author{Ken Shiozaki} 
\affiliation{Condensed Matter Theory Laboratory, RIKEN, Wako, Saitama 351-0198, Japan}

\author{Haruki Watanabe} 
\email[Correspondence should be addressed to H.W. at ]{haruki.watanabe@ap.t.u-tokyo.ac.jp}
\affiliation{Department of Applied Physics, University of Tokyo, Tokyo 113-8656, Japan}

\maketitle

\paragraph{Introduction and summary of the results.}---
Symmetry and topology are fundamentally related to each other. One key role of the symmetry is to protect and enrich the topological phases. For example, topological insulators and topological crystalline insulators are protected by the time-reversal symmetry and by space group symmetries~\cite{Z2_QSH, Joel_Z2, Z2_3D, RyuClassification, Kitaev, FuPRLTCI, FreedMoore, ChiuReflection, MorimotoClifford, BernevigPRL, Ken2014, LiuNonsymmorphic, Ken2016, HourGlass, Combinatorics, Ken2017}.  Another important role of the symmetry is to put constraints on the system and reduce allowed topological phases. For instance, Chern insulators do not exist when the time-reversal symmetry is assumed.  This competing effect of enrichment and reduction of topological phases makes the full classification of symmetry-protected topological phases intriguing and challenging.

There is another parallel relation between the symmetry and topology.  If one wants to examine the topological nature of quantum systems based on the very definition of the topological indices, one usually has to perform some sort of integrals. For example, the $\mathbb{Z}_2$-index of quantum spin Hall insulators is formulated as the integral of the so-called `Pfaffian' over the momentum space~\cite{Z2_QSH}.   The definition by itself might look simple, but the actual calculation requires a careful gauge-fixing and can be demanding~\cite{FuPump}. However, there is a shortcut with the help of inversion symmetry --- the celebrated Fu-Kane formula~\cite{Fu-Kane} allows one to determine the $\mathbb{Z}_2$-index just by multiplying the parity eigenvalues of the occupied bands over several high-symmetric points in the momentum space. This handy formula not merely significantly reduces the task but also gives us a practical guiding principle in material search --- it tells us that inducing a band inversion between two bands with opposite parities is sufficient to achieve a quantum spin Hall insulator.  The Fu-Kane formula is recently generalized to wider class of band topology and to more general class of spatial symmetries~\cite{NC, TopoChem, MSGnoninteracting, FangStudents}.

One of the achievements of this Letter is to establish a similar relation between the rotation symmetry and the quantized Hall conductance $\sigma_{xy}=\frac{e^2}{2\pi\hbar}C$ of many-body systems with arbitrary interactions. For example, for the two-fold rotation symmetry, it reads
\begin{equation}
e^{\pi i C}=w_{C_2}^{X}w_{C_2}^{Y}w_{C_2}^{\Gamma}w_{C_2}^{M},\label{C2int}
\end{equation}
where $w_{C_2}$ is the rotation eigenvalue of the many-body ground state.  Its noninteracting version was proven in Ref.~\onlinecite{Chern_Rotation} and the extension to interacting systems was hinted before~\cite{Chen}, but this problem has never been actually worked out so far.  To compute the many-body Chern number $C$ directly by definition~\cite{NTW,AvronSeiler}, one has to first find the many-body ground state as a smooth continuous function of the twisted angles of the boundary condition, and derive the Berry curvature by taking derivatives and finally perform the integral [see Eq.~\eqref{Chern} below].  Our formula can determine $C$ mod $n$ just by multiplying the eigenvalues of $n$-fold rotation $w_{C_n}$ at a few discrete values of the twisted angles.  

Apart from the symmetry, there is yet another key ingredient deeply related to the topology of many-body systems.  Given discrete translation symmetry and assuming the particle-number conservation, one can define the filling $\bar{\rho}$, the average number of particles per unit cell.  The (generalized) Lieb-Schultz-Mattis theorem~\cite{Oshikawa2000,Hastings2004,Sid2013,PNAS} tells us that gapped and symmetric ground states at a non-integral filling $\bar{\rho}\notin\mathbb{Z}$ must develop a ``topological order" accompanied by a fractionalization and topological degeneracy.  

Another interesting example of the interplay between the topology and the filling is in two dimensional periodic systems subjected to an external magnetic field of $2\pi p/q$-flux per unit cell.  In general, the many-body Chern number $C$ mod $q$ can be determined solely based on the filling $\bar{\rho}$ through
\begin{equation}
e^{2\pi i(\frac{p}{q}C-\bar{\rho})}=1.\label{classA}
\end{equation}
This relation was first derived for noninteracting band theory in Ref.~\onlinecite{DanaAvronZak} and later extended to interacting systems in Ref.~\onlinecite{Avron,Kol,YMLMO}.  Although the argument and conclusion of Ref.~\onlinecite{YMLMO} is quite intuitive, the actual derivation includes some mathematical subtleties in an essential manner.  For example, they made use of the time-evolution operator that adiabatically changes the flux piercing the torus as a part of `symmetries' of the Hamiltonian projected down to the ground state manifold.  The second result of the present Letter is to give an improved proof without such subtleties.  Eqs.~\eqref{C2int} and \eqref{classA} can be derived within the same framework in a completely parallel manner [compare Eqs.~\eqref{derivationEq1} and \eqref{derivationEq2}].

With this unified framework at hand, we finally explore a novel universal relation between the symmetry, topology, and filling. We find that Eqs.~\eqref{C2int} and \eqref{classA} can be combined into a new formula
\begin{equation}
e^{\pi i(\frac{p}{q}C-\bar{\rho})}=w_{C_2}^{\Gamma}w_{C_2}^{X}w_{C_2'}^{Y'}w_{C_2'}^{M'},\label{C2rotation}
\end{equation}
which can tell $C$ mod $2q$ in terms of the filling $\bar{\rho}$ and the rotation eigenvalues of many-body ground states.  See below Eq.~\eqref{derivationEq3} for the definition of $w_{C_2'}$.

We will justify Eqs.~\eqref{C2int}--\eqref{C2rotation} one by one in the reminder of the Letter. But before going into the technical derivation, let us develop an insight into Eq.~\eqref{C2int} through a much simpler example of a spin model.

\begin{figure}
\begin{center}
\includegraphics[width=1.0\columnwidth]{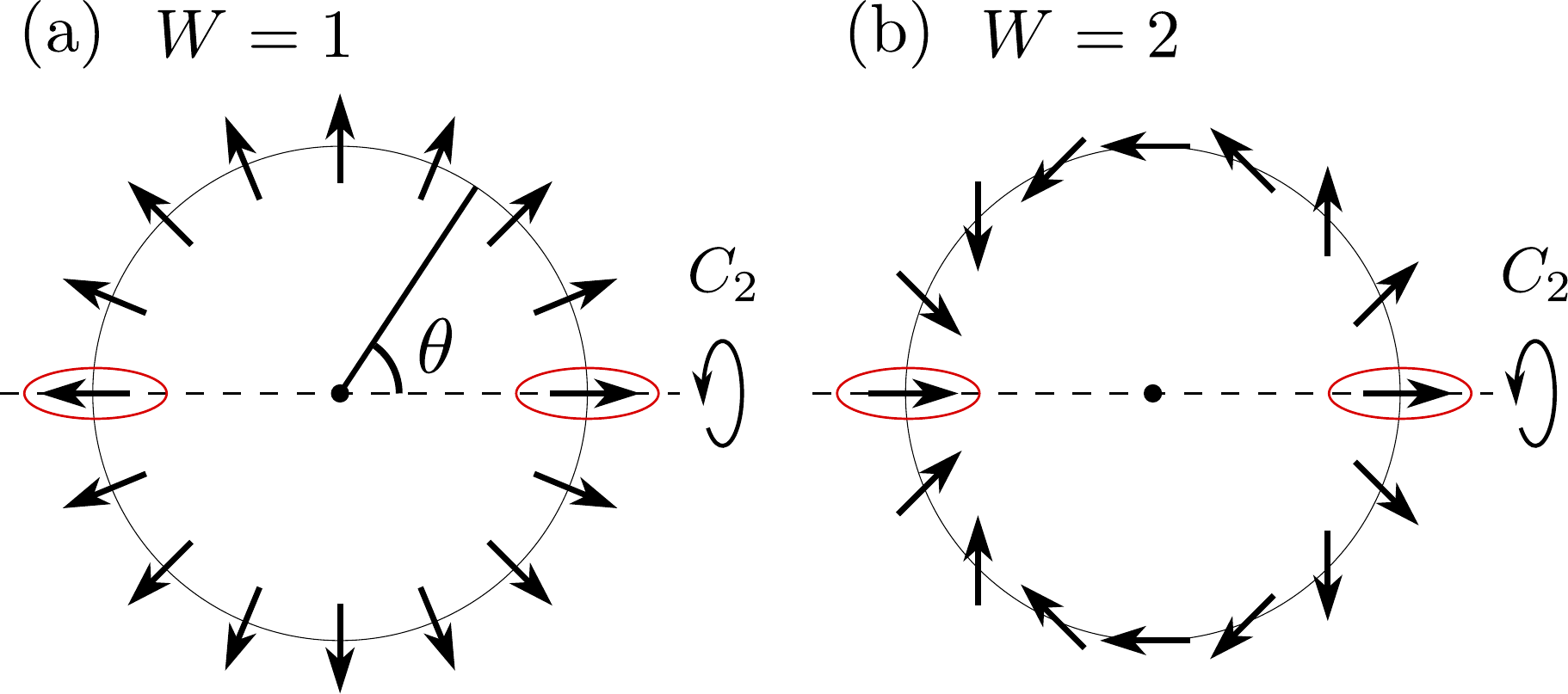}
\caption{The relation between the winding number $W\in\mathbb{Z}$ and the relative direction of the spins on the rotation axis. (a) $\vec{n}(0)\cdot\vec{n}(\pi)=-1$ and $W=+1$; (b) $\vec{n}(0)\cdot\vec{n}(\pi)=+1$ and $W=+2$. \label{fig:winding}}
\end{center}
\end{figure}

\paragraph{Winding number and symmetry.}---Consider classical spins $\vec{n}(\theta)$ ($0\leq\theta\leq2\pi$) on a 1D ring as illustrated in Fig.~\ref{fig:winding}. If spins are restricted into the $xy$-plane and if only smooth textures are allowed, the winding number $W[\vec{n}]=\frac{1}{2\pi}\int_0^{2\pi}d\theta\,\vec{n}(\theta)\times\partial_\theta\vec{n}(\theta)\in\mathbb{Z}$ of the map $\vec{n}: S^1\mapsto S^1$ is well-defined.  In principle one can actually perform this integral to find out the winding number.  Instead, here let us assume an additional $\pi$-rotation symmetry $C_2$ about the $x$ axis (the dashed line in Fig.~\ref{fig:winding}).  There are two special points on the ring, $\theta=0$ and $\pi$, left invariant under the rotation.  Under the symmetry, the spin can only point either $\pm\hat{x}$ at these points.  Clearly, there is a relation $e^{\pi i W[\vec{n}]}=\vec{n}(0)\cdot\vec{n}(\pi)$ between $W[\vec{n}]$ and $\vec{n}(0)\cdot\vec{n}(\pi)=\pm1$.  Hence, just by comparing the direction of the two spins in red circles in Fig.~\ref{fig:winding}, one can determine whether $W[\vec{n}]$ is even or odd.  This is a precise analog of Eq.~\eqref{C2int} --- limited information at high-symmetric points partially characterizes the topology.

\paragraph{Symmetries under twisted boundary condition.}---
Next, as preparation for our proof, let us summarize the symmetry of the Hamiltonian under a twisted boundary condition.  We start with a model defined on the infinitely-large square lattice $\vec{x}=(x,y)\in\mathbb{Z}^2$ with unit lattice constant~\footnote{More general settings, e.g., continuum models, can be treated in the same way}.  Suppose that the Hamiltonian $\hat{H}$ commutes with the particle number operator $\hat{N}\equiv\sum_{\vec{x}}\hat{n}_{\vec{x}}$ ($\hat{n}_{\vec{x}}\equiv\hat{c}_{\vec{x}}^\dagger \hat{c}_{\vec{x}}$), the translation operators $\hat{T}_x$ and $\hat{T}_y$, and the $\pi$-rotation $\hat{C}_2$ about the origin.  We put the system on the torus $T^2$ by introducing a twisted boundary condition 
\begin{equation}
(\hat{T}_x)^{L_x}=e^{-i\theta_x\hat{N}},\quad (\hat{T}_y)^{L_y}=e^{-i\theta_y\hat{N}},\label{twistedBC}
\end{equation}
identifying the annihilation operator $\hat{c}_{(L_x+1,y)}$ with $e^{i\theta_x}\hat{c}_{(1,y)}$, for example.  We express the resulting Hamiltonian $\hat{H}^{\vec{\theta}}$ [$\vec{\theta}=(\theta_x,\theta_y)$] in terms of the creation/annihilation operators in the range $x=1,2,\ldots,L_x$ and $y=1,2,\ldots,L_y$.   Note that $\hat{H}^{\vec{\theta}}$ has the period of $2\pi$ as a function of $\theta_{x}$ and $\theta_{y}$ as it is clear from the fact that the right-hand side of Eq.~\eqref{twistedBC} has that period.

\begin{figure}
\begin{center}
\includegraphics[width=1.0\columnwidth]{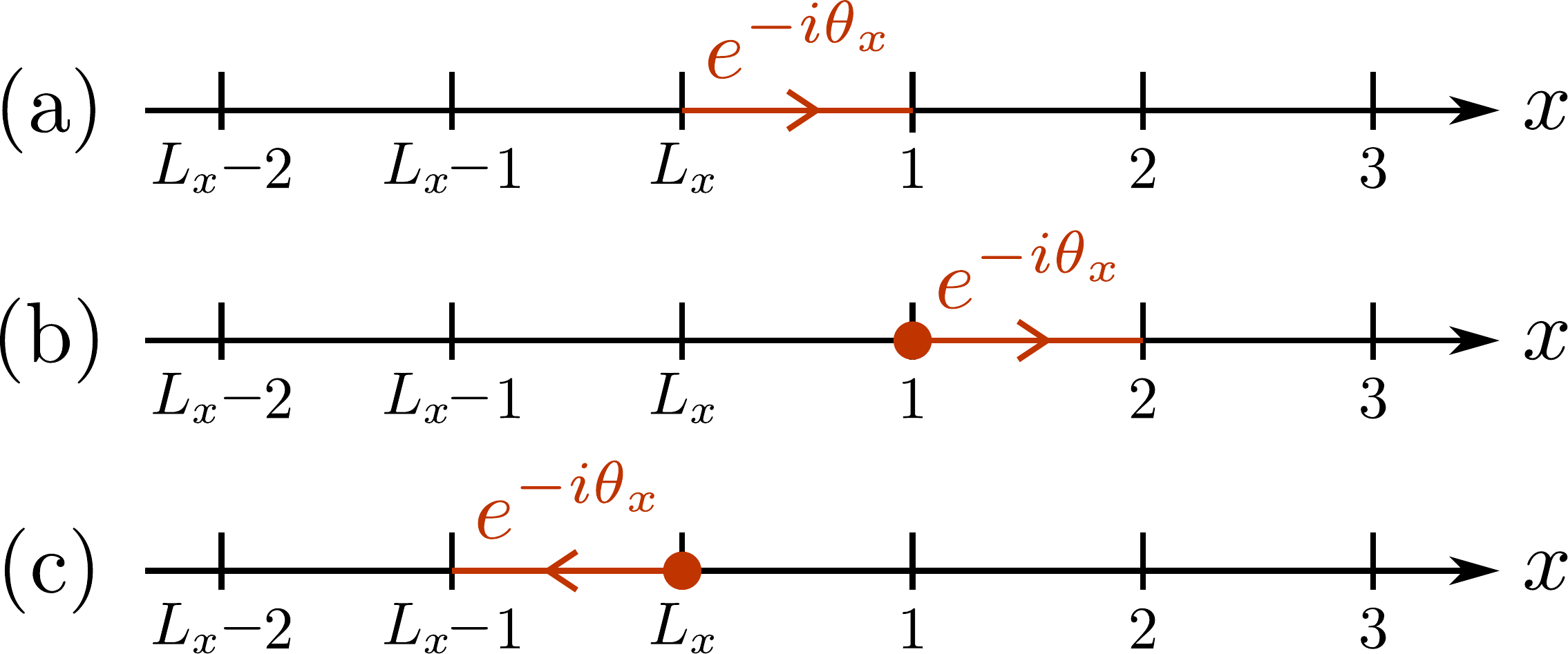}
\caption{Symmetries under twisted boundary condition in 1D. (a) The twisted boundary condition introduces the phase $e^{-i\theta_x}$ to the hopping from $x=L_x$ to $x=1$. (b) The bare translation moves the position of the twisted bond. (c) Similarly, the bare rotation about $x=L_x$ not only flips the phase but also moves the twisted bond.\label{fig1}}
\end{center}
\end{figure}

The twisted boundary condition introduces the phase $e^{-i\theta_x}$ to the hopping from $x=L_x$ to $x=1$ as shown in  Fig.~\ref{fig1}~(a).  Consequently, the bare translation is no longer a symmetry as it moves the position of the twisted hopping [see Fig.~\ref{fig1}~(b)]. To put the position back to the original one and leave the Hamiltonian unchanged, the translation should be followed by the local phase rotation $e^{-i\theta_x\hat{n}_{x=1}}$. Namely, the good translation symmetry commuting with $\hat{H}^{\vec{\theta}}$ is given by~\footnote{$\hat{T}_x$ and $\hat{T}_y$ in Eq.~\eqref{deftwistedTy} are the permutation operators defined by $\hat{T}_x\hat{c}_{(x,y)}\hat{T}_x^\dagger=\hat{c}_{(x+1,y)}$ for $x\leq L_x-1$ and $\hat{T}_x\hat{c}_{(L_x,y)}\hat{T}_x^\dagger=\hat{c}_{(1,y)}$.}
\begin{eqnarray}
\hat{T}_x^{\theta_x}\equiv e^{-i\theta_x\sum_{y}\hat{n}_{(1,y)}}\hat{T}_x,\quad
\hat{T}_y^{\theta_y}\equiv e^{-i\theta_y\sum_{x}\hat{n}_{(x,1)}}\hat{T}_y.\label{deftwistedTy}
\end{eqnarray}
The bare $\pi$-rotation also moves the position of the twisted hopping [Fig.~\ref{fig1}~(c)].  To recover the original position, the rotation should be accompanied by the local phase rotation~\footnote{$\hat{C}_2$ in Eq.~\eqref{C2HC1} is also the permutation operator defined by $\hat{C}_2\hat{c}_{(x,y)}\hat{C}_2^\dagger=\hat{c}_{(L_x-x,L_y-y)}$ ($x\leq L_x-1$ and $y\leq L_y-1$) and
$\hat{C}_2\hat{c}_{(x,L_y)}\hat{C}_2^\dagger=\hat{c}_{(L_x-x,L_y)}$ ($x\leq L_x-1$) and so on. For spinful particles,$\hat{C}_2$ should also include the spin rotation.}:
\begin{eqnarray}
&&\hat{C}_2^{\vec{\theta}}\equiv e^{-i\theta_x\sum_{y}\hat{n}_{(L_x,y)}-i\theta_y\sum_{x}\hat{n}_{(x,L_y)}}\hat{C}_2,\label{C2HC1}\\
&&\hat{C}_2^{\vec{\theta}}\hat{H}^{\vec{\theta}}(\hat{C}_2^{\vec{\theta}})^{\dagger}=\hat{H}^{-\vec{\theta}}.\label{C2HC2}
\end{eqnarray}
Observe that $\hat{C}_2^{\vec{\theta}}$ still flips the sign of the twisted angles, as suggested by the flipped arrow in Fig.~\ref{fig1}~(c). Hence, the Hamiltonian on torus lacks the rotation symmetry except for high-symmetric values of $\vec{\theta}$, i.e., $\Gamma=(0,0)$, $X=(\pi,0)$, $Y=(0,\pi)$, and $M=(\pi,\pi)$.  To fully make use of the rotation symmetry, one has to consider a parametric family of Hamiltonians $\hat{H}^{\vec{\theta}}$ as a function of $\vec{\theta}$.  

The above discussion suggests a similarity of the twisted angle $\vec{\theta}$ and the single-particle momentum $\vec{k}$ --- in fact Eq.~\eqref{twistedBC} implies $(k_x,k_y)\simeq(\theta_x/L_x,\theta_y/L_y)$. Therefore, $\vec{\theta}$ is the natural generalization of $\vec{k}$ in interacting systems.

\paragraph{Proof of Eq.~\eqref{C2int}.}---
Let us move on to the actual derivations.  Suppose that the ground state of $\hat{H}^{\vec{\theta}}$ is unique and gapped for all values of $\vec{\theta}$ and let $|\Phi^{\vec{\theta}}\rangle$ be the ground state.  The many-body Chern number $C$ is given in terms of the Berry connection $\vec{A}^{\vec{\theta}}=\langle\Phi^{\vec{\theta}}|\partial_{\vec{\theta}}|\Phi^{\vec{\theta}}\rangle$ as~\cite{NTW,AvronSeiler} (see also Supplemental Material A~\footnote{See our Supplemental Material, which includes Refs.~\cite{WatanabeOshikawa,NiuThouless}, for the review of many-body Chern number and symmetries under external fields, and for other expanded discussions.} ):
\begin{equation}
C=\frac{1}{2\pi i}\oint d^2\vec{\theta}\, F^{\vec{\theta}}, \quad F^{\vec{\theta}}=\nabla_{\vec{\theta}}\times\vec{A}^{\vec{\theta}}.\label{Chern}
\end{equation}

\begin{figure}
\begin{center}
\includegraphics[width=1.0\columnwidth]{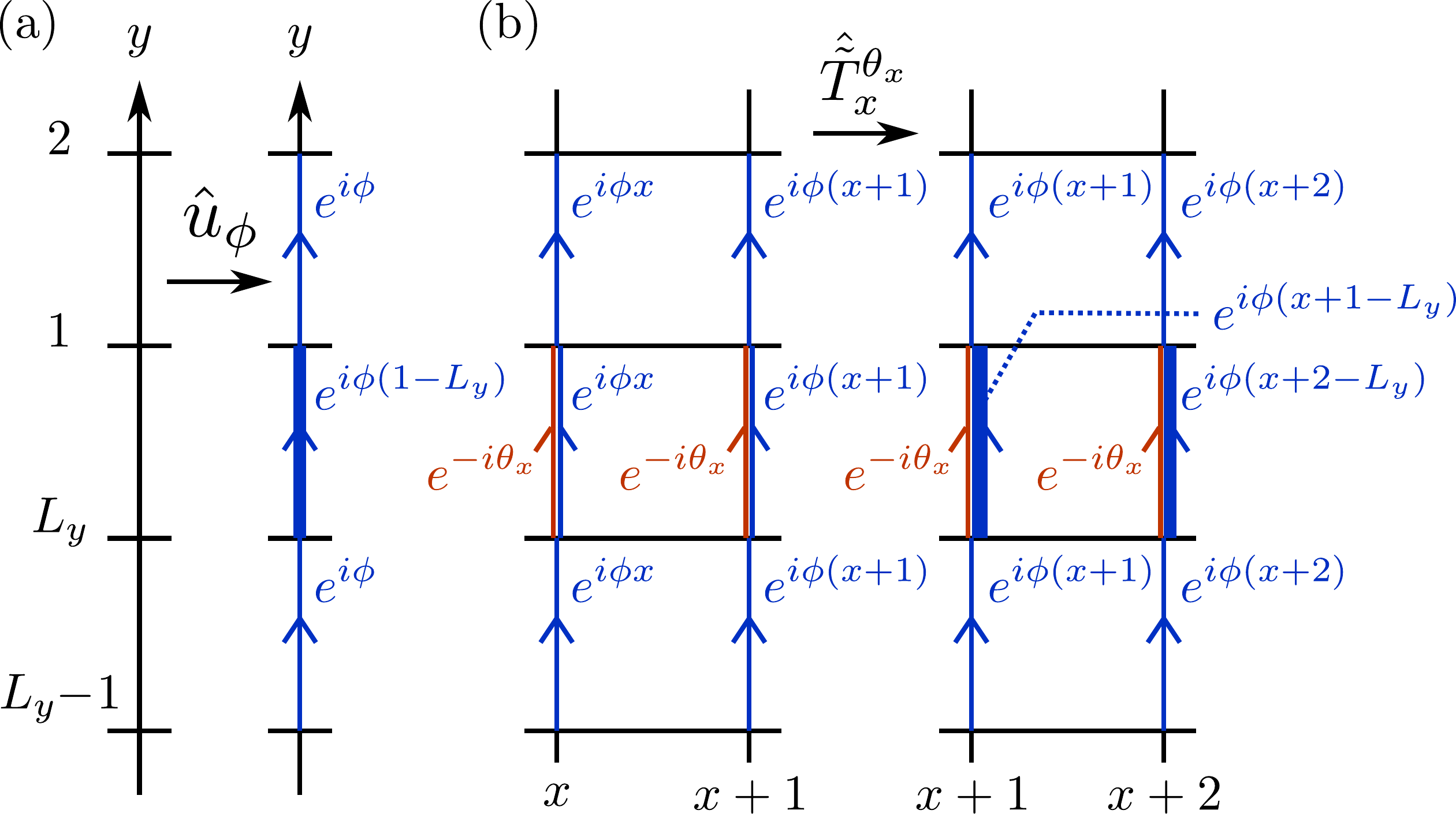}
\caption{(a) The action of $\hat{u}_{\phi}\equiv e^{i\phi\sum_{\vec{x}}y\hat{n}_{\vec{x}}}$ in $\hat{\tilde{T}}_x^{\theta_x}$ to the hopping in $y$ direction.  It multiplies a factor $e^{i\phi}$ to every hopping from $y$ to $y+1$ unless $y=L_y$. The hopping from $y=L_y$ to $1$ acquires the factor $e^{i\phi(1-L_y)}$.   (b) The action of $\hat{\tilde{T}}_x^{\theta_x}$ to the Hamiltonian $\hat{H}^{(\theta_x,\theta_y)}$. It effectively increases $\theta_y$ by $\phi L_y$\label{fig2}}
\end{center}
\end{figure}
As we assume the uniqueness of the ground state, Eq.~\eqref{C2HC2} suggests that $\hat{C}_2^{\vec{\theta}}|\Phi^{\vec{\theta}}\rangle$ is the ground state of $\hat{H}^{-\vec{\theta}}$. Hence, allowing for the phase ambiguity, we can write~\footnote{A nonzero Chern number puts an obstruction in choosing a global gauge. Thus Eqs.~\eqref{C2trans} and \eqref{AC22} should be understood as relations that hold only on one-dimensional circles in the $\bm{\theta}$ space.}
\begin{equation}
\hat{C}_2^{\vec{\theta}}|\Phi^{\vec{\theta}}\rangle=w_{C_2}^{\vec{\theta}}|\Phi^{-\vec{\theta}}\rangle\label{C2trans}.
\end{equation}
Consequently, the Berry connection at $\pm\vec{\theta}$ are related as
\begin{eqnarray}
&&\vec{A}^{-\vec{\theta}}=-\vec{A}^{\vec{\theta}}+i\vec{n}^{\vec{\theta}}+\nabla_{\vec{\theta}} \ln w_{C_2}^{\vec{\theta}},\label{AC22}\\
&&\vec{n}^{\vec{\theta}}\equiv i\langle\Phi^{\vec{\theta}}|(\hat{C}_2^{\vec{\theta}})^{\dagger}(\nabla_{\vec{\theta}}\hat{C}_2^{\vec{\theta}})|\Phi^{\vec{\theta}}\rangle.\label{AC2}
\end{eqnarray}
Using Eq.~\eqref{AC22} in Eq~\eqref{Chern}, we get
\begin{eqnarray}
&&\pi iC=\tfrac{1}{2}\oint d\theta_x \int_0^\pi d\theta_y[F^{\vec{\theta}}+F^{-\vec{\theta}}]\notag\\
&=&\oint d\theta_x \int_0^\pi d\theta_y[F^{\vec{\theta}}-\tfrac{i}{2}\nabla_{\vec{\theta}}\times \vec{n}^{\vec{\theta}}]\notag\\
&\overset{(\ast)}{=}&\oint d\theta_x [A_x^{(\theta_x,0)}-A_x^{(\theta_x,\pi)}-\tfrac{i}{2}n_x^{(\theta_x,0)}+\tfrac{i}{2}n_x^{(\theta_x,\pi)}]\notag\\
&=&\int_0^{\pi} d\theta_x [A_x^{(\theta_x,0)}+A_x^{(-\theta_x,0)}-in_x^{(\theta_x,0)}]\notag\\
&&-\int_0^{\pi} d\theta_x[A_x^{(\theta_x,\pi)}+A_x^{(-\theta_x,-\pi)}-in_x^{(\theta_x,\pi)}]\notag\\
&=&\int_0^{\pi} d\theta_x\partial_{\theta_x}\ln \frac{w_{C_2}^{(\theta_x,0)}}{w_{C_2}^{(\theta_x,\pi)}}=\ln\frac{w_{C_2}^{X}w_{C_2}^{Y}}{w_{C_2}^{\Gamma}w_{C_2}^{M}}.\label{derivationEq1}
\end{eqnarray}
Note that we used Stoke's theorem $\int_Sd^2\vec{\theta} F^{\vec{\theta}}=\oint_{\partial S}d\vec{\theta}\cdot\vec{A}^{\vec{\theta}}$ to go to the third line, which holds only when $\vec{A}^{\vec{\theta}}$ is smooth in $S$.  Singularities in $S$ give rise to corrections of $2\pi i\,m$ ($m\in\mathbb{Z}$). Thus, the step $(\ast)$ in Eq.~\eqref{derivationEq1} is true only modulo $2\pi i$.  In going to the fourth line, we used $n_x^{\vec{\theta}}=\textstyle\sum_{y}\langle\Phi^{\vec{\theta}}|\hat{n}_{(L_x,y)}|\Phi^{\vec{\theta}}\rangle=n_x^{-\vec{\theta}}$, which holds owing to the $\pi$-rotation symmetry.  

The last line of Eq.~\eqref{derivationEq1} contains $w_{C_2}^{K}$ at high-symmetric values of $\vec{\theta}=K$ invariant under $C_2$. According to Eq.~\eqref{C2trans}, $w_{C_2}^{K}$ in this case represents the $C_2$-eigenvalue of the many-body ground state $|\Phi^{K}\rangle$. Recalling that $(w_{C_2}^{K})^2=(-1)^{2SN}$ for the system of $N$-particles with spin $S$, we can rewrite \eqref{derivationEq1} as \eqref{C2int}.  Note that the proof never made use of the translation symmetry so that \eqref{C2int} applies, e.g., even in the presence of disorder.

\paragraph{Proof of Eq.~\eqref{classA}.}---
In order to discuss Eq.~\eqref{classA}, we have to introduce an external magnetic field of the strength $2\pi p/q$-flux per unit cell.  Here we assume $L_x$ an integer multiple of $q$ while $L_y$ co-prime with $q$.  In this setting, the translation should be modified to the magnetic translation
\begin{eqnarray}
\hat{\tilde{T}}_x^{\theta_x}\equiv e^{i\phi\sum_{\vec{x}}(y-\frac{L_y}{2})\hat{n}_{\vec{x}}}\hat{T}_x^{\theta_x},\label{deftwistedTx} 
\end{eqnarray}
while $\hat{T}_y^{\theta_y}$ and $\hat{C}_2^{\vec{\theta}}$ are unchanged (see Appendix B for a more general setting).  
A key observation here is that $\hat{\tilde{T}}_x^{\theta_x}$ shifts $\theta_y$ by $L_y\phi$ as described in Fig.~\ref{fig2}:
\begin{eqnarray}
\hat{\tilde{T}}_x^{\theta_x}\hat{H}^{\vec{\theta}}(\hat{\tilde{T}}_x^{\theta_x})^\dagger=\hat{H}^{\vec{\theta}+L_y\phi\hat{y}}.
\end{eqnarray}
Thus, indeed, $\hat{\tilde{T}}_x^{\theta_x}$ is not a symmetry of $\hat{H}^{\vec{\theta}}$ with a fixed $\vec{\theta}$.   This is reminiscent of the momentum shift in noninteracting systems under magnetic field, where $\hat{\tilde{T}}_x$ changes $k_y$ by $\phi$ due to the algebra $\hat{\tilde{T}}_y\hat{\tilde{T}}_x=e^{-i\phi\hat{N}}\hat{\tilde{T}}_x\hat{\tilde{T}}_y$.  
The Hamiltonian remains unchanged under $(\hat{\tilde{T}}_x^{\theta_x})^q$ as $qL_y\phi$ is an integer multiple of $2\pi$.

For brevity here we present the proof only for the simplest case of $\pi$-flux, i.e., $p=1$ and $q=2$. We include the proof for the most general case in Appendix C.  In this case $\hat{\tilde{T}}_x^{\theta_x}$ induces the shift $\theta_y\rightarrow\theta_y+\pi$.  Just as Eqs.~\eqref{C2trans}--\eqref{AC2}, we have
\begin{eqnarray}
&&\hat{\tilde{T}}_x^{\theta_x}|\Phi^{\vec{\theta}}\rangle=w_{T_x}^{\vec{\theta}}|\Phi^{\vec{\theta}+\pi\hat{y}}\rangle,\\
&&\vec{A}^{\vec{\theta}+\pi\hat{y}}=\vec{A}^{\vec{\theta}}-i\vec{t}^{\vec{\theta}}-\nabla_{\vec{\theta}} \ln w_{T_x}^{\vec{\theta}},\label{AT2}\\
&&\vec{t}^{\vec{\theta}}\equiv i\langle\Phi^{\vec{\theta}}|(\hat{\tilde{T}}_x^{\theta_x})^{\dagger}(\nabla_{\vec{\theta}}\hat{\tilde{T}}_x^{\theta_x})|\Phi^{\vec{\theta}}\rangle.
\end{eqnarray}
We can proceed in the same way as in Eq.~\eqref{derivationEq1} by replacing Eq.~\eqref{AC22} with Eq.~\eqref{AT2}:
\begin{eqnarray}
&&\pi iC=\tfrac{1}{2}\oint d\theta_x\int_{-\frac{\pi}{2}}^{\frac{\pi}{2}} d\theta_y[F^{\vec{\theta}}+F^{\vec{\theta}+\pi\hat{y}}]\notag\\
&=&\oint d\theta_x\int_{-\frac{\pi}{2}}^{\frac{\pi}{2}}  d\theta_y[F^{\vec{\theta}}-\tfrac{i}{2}\nabla_{\vec{\theta}}\times \vec{t}^{\vec{\theta}}]\notag\\
&\overset{(\ast)}{=}&\oint d\theta_x [A_x^{(\theta_x,-\frac{\pi}{2})}-A_x^{(\theta_x,\frac{\pi}{2})}-\tfrac{i}{2}t_x^{(\theta_x,-\frac{\pi}{2})}+\tfrac{i}{2}t_x^{(\theta_x,\frac{\pi}{2})}]\notag\\
&=&\oint d\theta_x [\partial_{\theta_x} \ln w_{T_x}^{(\theta_x,-\frac{\pi}{2})}+\tfrac{i}{2}t_x^{(\theta_x,-\frac{\pi}{2})}+\tfrac{i}{2}t_x^{(\theta_x,\frac{\pi}{2})}]\notag\\
&=&\oint d\theta_x i\bar{\rho}L_y=2\pi i\bar{\rho}L_y.\label{derivationEq2}
\end{eqnarray}
Again the step $(\ast)$ is true only modulo $2\pi i$. In going to the last line, we used $
t_x^{\vec{\theta}}=n_x^{\vec{\theta}}$ and $n_x^{\vec{\theta}}+n_x^{\vec{\theta}+\pi\hat{y}}=2\bar{\rho}L_y$, 
which follow by the (magnetic) translation symmetry.  Here, $\bar{\rho}$ is the average number of particles in the original (i.e. not-enlarged) unit cell.
The Lieb-Schultz-Mattis theorem~\cite{Oshikawa2000,Hastings2004,Sid2013,PNAS} demands that $q\bar{\rho}$, the filling with respect to the right unit cell of $\hat{H}^{\vec{\theta}}$, be an integer to be consistent with the assumption of the unique and gapped ground state. Hence, $e^{2\pi i\bar{\rho}L_y}=e^{2\pi i\bar{\rho}}$ and we arrive at Eq.~\eqref{classA}.

\paragraph{Proof of Eq.~\eqref{C2rotation}.}---
We have verified Eqs.~\eqref{C2int} and \eqref{classA}, which determine the many-body Chern number $C$ mod $2$ and $q$, respectively, based on the many-body rotation eigenvalues and the filling.  When $q$ is odd, one can use these formulas separately and compute $C$ mod $2q$.  This is not the case when $q$ is even.  Our new formula in Eq.~\eqref{C2rotation} goes beyond this naive combination as it works even when $q$ is even.  Let us now present the proof, again for $q=2$. 

Starting from the second line of Eq.~\eqref{derivationEq2}, we have
\begin{eqnarray}
&&\tfrac{1}{2}\pi iC=\tfrac{1}{2}\oint d\theta_x\int_{-\frac{\pi}{2}}^{\frac{\pi}{2}}  d\theta_y[F^{\vec{\theta}}-\tfrac{i}{2}\nabla_{\vec{\theta}}\times \vec{t}^{\vec{\theta}}]\notag\\
&=&\tfrac{1}{2}\oint d\theta_x\int_{0}^{\frac{\pi}{2}}  d\theta_y[F^{\vec{\theta}}+F^{-\vec{\theta}}-\tfrac{i}{2}\nabla_{\vec{\theta}}\times (\vec{t}^{\vec{\theta}}-\vec{t}^{-\vec{\theta}})]\notag\\
&=&\oint d\theta_x\int_{0}^{\frac{\pi}{2}}  d\theta_y[F^{\vec{\theta}}-\tfrac{i}{2}\nabla_{\vec{\theta}}\times \vec{n}^{\vec{\theta}}]\notag\\
&\overset{(\ast)}{=}&\oint d\theta_x [A_x^{(\theta_x,0)}-A_x^{(\theta_x,\frac{\pi}{2})}-\tfrac{i}{2}n_x^{(\theta_x,0)}+\tfrac{i}{2}n_x^{(\theta_x,\frac{\pi}{2})}]\notag\\
&=&\int_0^{\pi} d\theta_x [A_x^{(\theta_x,0)}+A_x^{(-\theta_x,0)}-in_x^{(\theta_x,0)}]\notag\\
&&-\int_0^{\pi} d\theta_x[A_x^{(\theta_x,\frac{\pi}{2})}+A_x^{(-\theta_x,-\frac{\pi}{2})}-in_x^{(\theta_x,\frac{\pi}{2})}]\notag\\
&&-\int_0^{\pi} d\theta_x[A_x^{(-\theta_x,\frac{\pi}{2})}-A_x^{(-\theta_x,-\frac{\pi}{2})}+in_x^{(-\theta_x,-\frac{\pi}{2})}]\notag\\
&&+\tfrac{i}{2}\int_0^{\pi} d\theta_x[n_x^{(\theta_x,\frac{\pi}{2})}+n_x^{(-\theta_x,\frac{\pi}{2})}]\notag\\
&=&\ln \frac{w_{C_2}^Xw_{C_2'}^{(0,\frac{\pi}{2})}}{w_{C_2}^\Gamma w_{C_2'}^{(\pi,\frac{\pi}{2})}}+\pi i\bar{\rho}L_y,\label{derivationEq3}
\end{eqnarray}
where $w_{C_2'}^{\vec{\theta}}\equiv w_{T_x}^{-\vec{\theta}}w_{C_2}^{\vec{\theta}}$ is the phase factor for the product $\hat{C}_2^{'\vec{\theta}}\equiv\hat{\tilde{T}}_x^{-\theta_x}\hat{C}_2^{\vec{\theta}}$. When $L_y=4\ell+1$ ($\ell\in\mathbb{Z}$), Eq.~\eqref{derivationEq3} is precisely \eqref{C2rotation} if one defines $Y'=(0,\frac{\pi}{2})$ and $M'=(\pi,\frac{\pi}{2})$.  When $L_y=4\ell-1$, one has to instead look at $Y'=(0,-\frac{\pi}{2})$ and $M'=(\pi,-\frac{\pi}{2})$ to get Eq.~\eqref{C2rotation}.

\paragraph{Discussion and outlook.}---
So far we have focused only on the two-fold rotation. Here let us discuss the possibility of extending Eqs.~\eqref{C2int} and \eqref{C2rotation} to higher-order rotations.  We can derive the formula corresponding to Eq.~\eqref{C2int}, for three, four, and six-fold rotation symmetry, which respectively reads
\begin{eqnarray}
e^{\frac{-2\pi i C}{3}}=(-1)^{2SN}w_{C_3}^{\Gamma}w_{C_3}^{K}w_{C_3}^{K'},\label{C3int}\\
e^{\frac{-2\pi i C}{4}}=(-1)^{2SN}w_{C_4}^{\Gamma}w_{C_4}^{M}w_{C_2}^{X},\label{C4int}\\
e^{\frac{-2\pi i C}{6}}=(-1)^{2SN}w_{C_6}^{\Gamma}w_{C_3}^{K}w_{C_2}^{M},\label{C6int}
\end{eqnarray}
See Appendix~D for the details. In contrast, we have not succeeded in deriving the higher-order rotation version of Eq.~\eqref{C2rotation} in general.  The crucial difference between the two-fold rotation and other rotations lies in the symmetry algebra. That is, translations in two different directions are related by $C_n$ ($n=3,4,6$), e.g., $\hat{C}_4\hat{T}_x\hat{C}_4^{-1}=\hat{T}_y$, while it is not the case for $C_2$.  As a result, one cannot choose $L_x$ and $L_y$ independently to be consistent with $C_n$ ($n=3,4,6$) symmetry.  At this moment it is not clear how one can deal with a general $\phi=2\pi p/q$ flux together with $C_4$ rotation symmetry when $q$ is even. However, case-by-case studies for the $\pi$-flux and the $\frac{\pi}{2}$-flux reveals that there is a way to handle specific values of $\phi$ with $C_4$. We derive formulas that tell $C$ mod $4q$ for these cases in Appendix~E.  

The formalism developed in this work can also be applied to fractional quantum Hall states by relaxing the assumption of the uniqueness of the ground state on the torus.  The $D$-fold degenerate ground states interchange among them when $\theta_x$ is increased by $2\pi n$ $(n=1,2,\cdots D-1)$, and the formula for the many-body Chern number should be modified to $\tilde{C}=\frac{1}{2\pi i D}\int_{0}^{2\pi D}  d\theta_x\int_{0}^{2\pi} d\theta_y\,F^{\vec{\theta}}$, which is related to the quantized Hall conductivity as $\sigma_{xy}=\frac{e^2}{2\pi \hbar}\tilde{C}$.  In Appendix~F, we derive 
\begin{equation}
e^{2\pi i\left(\frac{p}{q}\tilde{C}-\bar{\rho}\right)D}=1
\end{equation}
in this setting. One should also be able to incorporate with rotation symmetries.

When the magnetic flux $\phi$ per unit cell equals $\pi$, the system may recover the time-reversal symmetry.  In such a case one can explore the relation between the (many-body) $\mathbb{Z}_2$ quantum spin Hall index and the filling~\cite{YML} and the rotation eigenvalues. We leave these interesting extensions to future work.\\

\begin{acknowledgments}
H.W. deeply thanks Chen Fang for illuminating discussion on the relation between the many-body Chern number and the rotation eigenvalues.  
H.W. also thanks Ashvin Vishwanath and Hoi Chun Po for discussions on a related project.
The work of H.W. is supported by JSPS KAKENHI Grant Number JP17K17678.  K.S. is supported by RIKEN Special Postdoctoral Researcher Program.  
\end{acknowledgments}

\bibliography{references}
\clearpage

\appendix

\onecolumngrid

\section{Appendix A: The definition of the many-body Chern number}
In this appendix, we clarify the physical meaning of the many-body Chern number.  In the main text, we defined the Chern number as
\begin{equation}
C_{\text{b}}=\frac{1}{2\pi i}\oint d^2\vec{\theta}\, F_{\text{b}}^{\vec{\theta}}, \quad F_{\text{b}}^{\vec{\theta}}=\nabla_{\vec{\theta}}\times\vec{A}_{\text{b}}^{\vec{\theta}},\quad \vec{A}_{\text{b}}^{\vec{\theta}}=\langle\Phi_{\text{b}}^{\vec{\theta}}|\partial_{\vec{\theta}}|\Phi_{\text{b}}^{\vec{\theta}}\rangle.
\end{equation}
Here, $\vec{\theta}=(\theta_x, \theta_y)$ is the twisted angle of the boundary condition and $|\Phi_{\text{b}}^{\vec{\theta}}\rangle$ is the ground state of $\hat{H}_{\text{b}}^{\vec{\theta}}$. In order to clarify that only the hopping near the boundary is twisted, here we added the subscript `b'.  Alternatively, by the unitary transformation
\begin{eqnarray}
\hat{H}_{\text{u}}^{\vec{\theta}}&\equiv&e^{-i\vec{\theta}\cdot\vec{\hat{P}}}\hat{H}_{\text{b}}^{\vec{\theta}}e^{i\vec{\theta}\cdot\vec{\hat{P}}}\\
|\Phi_{\text{u}}^{\vec{\theta}}\rangle&\equiv&e^{-i\vec{\theta}\cdot\vec{\hat{P}}}|\Phi_{\text{b}}^{\vec{\theta}}\rangle,\\
\hat{\vec{P}}&\equiv&\left(\sum_{\vec{x}}\frac{x}{L_x}\hat{n}_{\vec{x}},\sum_{\vec{x}}\frac{y}{L_y}\hat{n}_{\vec{x}}\right),
\end{eqnarray}
we can spread the effect of twisted boundary condition uniformly to the entire space. As a result, $\hat{H}_{\text{u}}^{\vec{\theta}}$ possesses the bare translation $\hat{T}$.   One can define another Chern number using $|\Phi_{\text{u}}^{\vec{\theta}}\rangle$:
\begin{equation}
C_{\text{u}}=\frac{1}{2\pi i}\oint d^2\vec{\theta}\, F_{\text{u}}^{\vec{\theta}}, \quad F_{\text{u}}^{\vec{\theta}}=\nabla_{\vec{\theta}}\times\vec{A}_{\text{u}}^{\vec{\theta}},\quad \vec{A}_{\text{u}}^{\vec{\theta}}=\langle\Phi_{\text{u}}^{\vec{\theta}}|\partial_{\vec{\theta}}|\Phi_{\text{u}}^{\vec{\theta}}\rangle.
\end{equation}
This is, in fact, the original definition introduced by Niu, Thouless, and Wu~\cite{NTW}.  One may wonder which of the two Chern numbers corresponds to the Hall conductance, but we can easily show $C_{\text{b}}=C_{\text{u}}$~\cite{WatanabeOshikawa}.
\begin{eqnarray}
F_{\text{u}}^{\vec{\theta}}-F_{\text{b}}^{\vec{\theta}}=-i\nabla_{\vec{\theta}}\times\langle\Phi_{\text{b}}^{\vec{\theta}}|\vec{\hat{P}}|\Phi_{\text{b}}^{\vec{\theta}}\rangle.
\end{eqnarray}
Since $\langle\Phi_{\text{b}}^{\vec{\theta}}|\vec{\hat{P}}|\Phi_{\text{b}}^{\vec{\theta}}\rangle$ is a periodic function of $\theta_{x,y}$ with the period $2\pi$, this total derivative term does not contribute to the integral.   

The physical meaning of $C\equiv C_{\text{u}}=C_{\text{b}}$ is provided by Laughlin-type argument~\cite{NTW}.  Let us define $\mathcal{P}_x(\theta_y)\equiv i\int_0^{2\pi}\tfrac{d\theta_x}{2\pi}\,A_{\text{u},x}^{\vec{\theta}}=\int_0^{2\pi}\tfrac{d\theta_x}{2\pi}\,i\langle \Phi_{\text{u}}^{\vec{\theta}}|\partial_{\theta_x}|\Phi_{\text{u}}^{\vec{\theta}}\rangle$ and $\mathcal{P}_y(\theta_x)\equiv i\int_0^{2\pi}\tfrac{d\theta_y}{2\pi}\,A_{\text{u},y}^{\vec{\theta}}=\int_0^{2\pi}\tfrac{d\theta_y}{2\pi}\,i\langle \Phi_{\text{u}}^{\vec{\theta}}|\partial_{\theta_y}|\Phi_{\text{u}}^{\vec{\theta}}\rangle$. If we choose the gauge periodic in $\theta_x$, then we can rewrite $C$ as
\begin{eqnarray}
C=\int_0^{2\pi}d\theta_y\,\partial_{\theta_y}\mathcal{P}_x(\theta_y)-\int_0^{2\pi}d\theta_x\,\partial_{\theta_x}\mathcal{P}_y(\theta_x)=\int_0^{2\pi}d\theta_y\,\partial_{\theta_y}\mathcal{P}_x(\theta_y)=\mathcal{P}_x(2\pi)-\mathcal{P}_x(0).
\end{eqnarray}
Suppose that $\theta_y=\theta_y(t)$ has a weak time-dependence, increasing from $0$ to $2\pi$.  On the one hand, Faraday's law tells us that there will be an induced electric field $E_y=\frac{\partial_t\theta_y(t)}{L_y}$.  The transported charge $Q$ during this period is related to $\sigma_{xy}$ as
\begin{eqnarray}
Q=\int_0^TdtJ_x(t)=L_y\sigma_{xy}\int_0^TdtE_y(t)=\sigma_{xy}\int_0^Tdt\partial_t\theta_y(t)=2\pi\sigma_{xy}.
\end{eqnarray}
On the other hand, from the Thouless-pump point of view~\cite{NiuThouless}, we have 
\begin{eqnarray}
Q=\int_{0}^Tdt\partial_t\mathcal{P}_x(\theta_y(t))=\mathcal{P}_x(2\pi)-\mathcal{P}_x(0).
\end{eqnarray}
Therefore,
\begin{equation}
C=Q=2\pi\sigma_{xy}.
\end{equation}

\clearpage
\section{Appendix B: Symmetries under magnetic field}
\label{app:Symmetries}
In this appendix we summarize symmetries under uniform magnetic field and twisted boundary conditions in a general setting.   Here we do not restrict $x,y$ to be integers, and we do not assume $L_x$ is an integer multiple of $q$. We will only assume $(p/q) L_xL_y$ is an integer.

\subsection{Landau gauge}
In the Landau gauge, twisted translations and rotations read
\begin{eqnarray}
\hat{\tilde{T}}_x^{\vec{\theta}}&=& \hat{T}_xe^{-i\sum_{y=1}^{L_y}[\theta_x+\phi L_x(y-\frac{L_y}{2})]\hat{n}_{(L_x,y)}}e^{i\phi\sum_{\vec{x}}(y-\frac{L_y}{2})\hat{n}_{\vec{x}}},\\
\hat{\tilde{T}}_y^{\vec{\theta}}&=& \hat{T}_ye^{-i\sum_{x=1}^{L_x}\theta_y\hat{n}_{(x,L_y)}},\\
\hat{C}_2^{\vec{\theta}}&=&\hat{C}_2e^{-i\sum_{y=1}^{L_y}[\theta_x+\phi L_x(y-\frac{L_y}{2})]\hat{n}_{(L_x,y)}-i\sum_{x=1}^{L_x}\theta_y\hat{n}_{(x,L_y)}},\\
\hat{C}_4^{\vec{\theta}}&=&\hat{C}_4e^{-i\phi \sum_{\vec{x}}(x-\frac{L_x}{2})(y-\frac{L_y}{2})\hat{n}_{\vec{x}}}e^{-i\sum_{x=1}^{L_x}[\theta_y-\phi L_y(x-\frac{L_x}{2})]\hat{n}_{(x,L_y)}}.
\end{eqnarray}

\subsection{Symmetric gauge}
In the symmetric gauge, they are
\begin{eqnarray}
\hat{\tilde{T}}_x^{\theta_x}&=& \hat{P}_xe^{-i\sum_{y=1}^{L_y}[\theta_x+\frac{\phi}{2}L_x(y-\frac{L_y}{2})]\hat{n}_{(L_x,y)}}e^{\frac{i\phi}{2}\sum_{\vec{x}}(y-\frac{L_y}{2})\hat{n}_{\vec{x}}},\\
\hat{\tilde{T}}_y^{\theta_y}&=& \hat{P}_ye^{-i\sum_{x=1}^{L_x}[\theta_y-\frac{\phi}{2}L_y(x-\frac{L_x}{2})]\hat{n}_{(x,L_y)}}e^{-\frac{i\phi}{2}\sum_{\vec{x}}(x-\frac{L_x}{2})\hat{n}_{\vec{x}}},\\
\hat{C}_2^{\vec{\theta}}&=&\hat{P}_2e^{-i\sum_{y=1}^{L_y}[\theta_x+\frac{\phi}{2}  L_x(y-\frac{L_y}{2})]\hat{n}_{(L_x,y)}-i\sum_{x=1}^{L_x}[\theta_y-\frac{\phi}{2} L_y(x-\frac{L_x}{2})]\hat{n}_{(x,L_y)}}e^{i\frac{\phi}{2}L_xL_y\hat{n}_{(L_x,L_y)}},\\
\hat{C}_4^{\vec{\theta}}&=&\hat{P}_4e^{-i\sum_{x=1}^{L_x}[\theta_y-\frac{\phi}{2} L_y(x-\frac{L_x}{2})]\hat{n}_{(x,L_y)}}.
\end{eqnarray}

\subsection{Commutation relations}
Operations above satisfy the following algebra
\begin{eqnarray}
&&(\hat{\tilde{T}}_x^{\vec{\theta}})^{L_x}=e^{-i\theta_x\hat{N}},\\
&&(\hat{\tilde{T}}_y^{\vec{\theta}})^{L_y}=e^{-i\theta_y\hat{N}},\\
&&\hat{\tilde{T}}_x^{\vec{\theta}-L_x\phi\hat{x}}\hat{\tilde{T}}_y^{\vec{\theta}}=e^{i\phi\hat{N}}\hat{\tilde{T}}_y^{\vec{\theta}+L_y\phi\hat{y}}\hat{\tilde{T}}_x^{\vec{\theta}},\\
&&\hat{C}_2^{-\vec{\theta}}\hat{C}_2^{\vec{\theta}}=1,\\
&&\hat{C}_2^{\vec{\theta}+L_y\phi\hat{y}}\hat{\tilde{T}}_x^{\vec{\theta}}=(\hat{\tilde{T}}_x^{-\vec{\theta}})^{-1}\hat{C}_2^{\vec{\theta}},\\
&&\hat{C}_2^{\vec{\theta}-L_x\phi\hat{x}}\hat{\tilde{T}}_y^{\vec{\theta}}=(\hat{\tilde{T}}_y^{-\vec{\theta}})^{-1}\hat{C}_2^{\vec{\theta}},\\
&&\hat{C}_4^{(-\theta_y,\theta_x)}\hat{C}_4^{\vec{\theta}}=\hat{C}_2^{\vec{\theta}},\\
&&\hat{C}_4^{\vec{\theta}+L_y\phi\hat{y}}\hat{\tilde{T}}_x^{\vec{\theta}}=\hat{\tilde{T}}_y^{(-\theta_y,\theta_x)}\hat{C}_4^{\vec{\theta}},\\
&&\hat{C}_4^{\vec{\theta}-L_x\phi\hat{x}}\hat{\tilde{T}}_y^{\vec{\theta}}=(\hat{\tilde{T}}_x^{(-\theta_y,\theta_x)})^{-1}\hat{C}_4^{\vec{\theta}}.
\end{eqnarray}
To verify these relations one has to use the operator identity $e^{i\phi L_xL_y\hat{n}_{\vec{x}}}=1$. Also, those including $\hat{C}_4$ are valid only when $L_x=L_y$.

\clearpage

\section{Appendix C: Proof of Eqs.~(2) and (3) for a general flux $2\pi p/q$}

In this appendix, we will prove Eqs.~(2) and (3) in the main text in the most general setting.
\subsection{proof of Eq.~(2)}

As discussed in the main text, $\hat{\tilde{T}}_x^{\theta_x}$ shifts $\theta_y$ by $d\equiv L_y\phi$.  Since we assume that the ground state is unique and gapped for the all values of $\theta_x$ and $\theta_y$, we can write
\begin{equation}
(\hat{\tilde{T}}_x^{\theta_x})^m\ket{\Phi^{\bm{\theta}}}=w^{\bm{\theta}}_{T_x^m}\ket{\Phi^{\bm{\theta}+md\hat{y}}}.
\end{equation}
As a consequence the Berry connection at $\vec{\theta}$ and $\vec{\theta}+md\hat{y}$ are related as
\begin{equation}
\bm{A}^{\bm{\theta}+md\hat{y}}=\bm{A}^{\bm{\theta}}-i\sum_{j=0}^{m-1}\bm{t}^{\bm{\theta}+jd\hat{y}}-\nabla_{\bm{\theta}}\ln w^{\bm{\theta}}_{T_x^m},\quad \vec{t}^{\vec{\theta}}\equiv (n_x^{\vec{\theta}},0),\quad n_x^{\vec{\theta}}=\sum_y\langle\Phi^{\vec{\theta}}|\hat{n}_{(L_x,y)}|\Phi^{\vec{\theta}}\rangle.\label{sB1}
\end{equation}
Because of the magnetic translation symmetry $(\hat{\tilde{T}}_x^{\theta_x})^q$, $n_x^{\vec{\theta}}$ satisfies
\begin{equation}
\sum_{m=0}^{q-1}n_x^{\bm{\theta}+md\hat{y}}=qL_y\bar{\rho}.
\end{equation}

Let us now evaluate the Chern number by reducing the irreducible part by a factor of $1/q$ using $\hat{\tilde{T}}_x^{\theta_x}$.  
To this end, recall that the shift of $\theta_y$ induced by $\hat{\tilde{T}}_x^{\theta_x}$,  $d=L_y\phi$, is not small at all; its proportional to $L_y$.  We can consider ``$d$ mod $2\pi$" as is done in the main text, but we find that quantity may not always be the easiest to deal with. Here, instead, we enlarge the integration range of $\theta_y$ by $\frac{qd}{2\pi}=\frac{qL_y\phi}{2\pi}=pL_y$ times and evaluate $p L_y C$ rather than $C$ itself:
\begin{eqnarray}
2\pi ip L_y C&=&pL_y\oint d\theta_x\oint d\theta_y\,F^{\bm{\theta}}=\oint d\theta_x\int_{-\frac{d}{2}}^{-\frac{d}{2}+2\pi pL_y} d\theta_y\,F^{\bm{\theta}}=\sum_{m=0}^{q-1}\oint d\theta_x\int_{-\frac{d}{2}}^{-\frac{d}{2}} d\theta_y\,F^{\bm{\theta}+md\hat{y}}.
\end{eqnarray}
As we are only interested in $C$ mod $q$ and $pL_y$ and $q$ are co-prime, we lose nothing by this manipulation. Plugging Eq.~\eqref{sB1}, we get
\begin{eqnarray}
2\pi ip L_y C&=&\sum_{m=0}^{q-1}\oint d\theta_x\int_{-\frac{d}{2}}^{-\frac{d}{2}} d\theta_y\,F^{\bm{\theta}+md\hat{y}}=q\oint d\theta_x\int_{-\frac{d}{2}}^{\frac{d}{2}}  d\theta_y\,F^{\bm{\theta}}-i\sum_{m=1}^{q-1}\sum_{j=0}^{m-1}\oint d\theta_x\int^{\frac{d}{2}}_{-\frac{d}{2}}d\theta_y\,\nabla_{\bm{\theta}}\times\bm{t}^{\bm{\theta}+jd\hat{y}}\notag\\
&=&q\oint d\theta_x\int_{-\frac{d}{2}}^{\frac{d}{2}}  d\theta_y\,F^{\bm{\theta}}-i\sum_{m=1}^{q-1}\sum_{j=0}^{m-1}\oint d\theta_x\,\left[n_x^{(\theta_x,-\frac{d}{2}+jd)}-n_x^{(\theta_x,-\frac{d}{2}+(j+1)d)}\right]\notag\\
&\overset{(\ast)}{=}&q\oint d\theta_x\,[A_x^{(\theta_x,-\frac{d}{2})}-A_x^{(\theta_x,\frac{d}{2})}]-i\sum_{m=1}^{q-1}\sum_{j=0}^{m-1}\oint d\theta_x\,\left[n_x^{(\theta_x,-\frac{d}{2}+jd)}-n_x^{(\theta_x,-\frac{d}{2}+(j+1)d)}\right]\notag\\
&=&\left(q\oint d\theta_x \,\partial_{\theta_x}\ln w^{(\theta_x,-\frac{d}{2})}_{T_x}+iq\oint d\theta_x\,n_x^{(\theta_x,-\frac{d}{2})}\right)-\left(iq\oint d\theta_x\,n_x^{(\theta_x,-\frac{d}{2})}-2\pi iqL_y\bar{\rho}\right)\notag\\
&=&2\pi iqL_y\bar{\rho}.\label{sB2}
\end{eqnarray}
In going to the second last line, we used Eq.~\eqref{sB1} to the first term and 
\begin{eqnarray}
i\sum_{m=1}^{q-1}\sum_{j=0}^{m-1}\oint d\theta_x\,\left[n_x^{(\theta_x,-\frac{d}{2}+jd)}-n_x^{(\theta_x,-\frac{d}{2}+(j+1)d)}\right]&=&iq\oint d\theta_x\,n_x^{(\theta_x,-\frac{d}{2})}-i\sum_{m=0}^{q-1}\oint d\theta_x\,n_x^{(\theta_x,-\frac{d}{2}+md)}\notag\\
&=&iq\oint d\theta_x\,n_x^{(\theta_x,-\frac{d}{2})}-2\pi i qL_y\bar{\rho}\label{sB22}
\end{eqnarray}
to the second term. Also, in the last line we dropped $\oint d\theta_x\,\partial_{\theta_x} \ln w_{T_x}^{(\theta_x,-\frac{d}{2})}$ based on the periodicity of $w_{T_x}^{(\theta_x,-\frac{d}{2})}$.  The step $(\ast)$ in Eq.~(\ref{sB2}) is true only modulo $2\pi iq$. Therefore, we have 
\begin{equation}
2\pi ip L_y C=2\pi iqL_y\bar{\rho}\quad\text{mod}\quad 2\pi i q.
\end{equation}
Since $L_y$ is co-prime with $q$, we can rewrite $2\pi i(\frac{p}{q}C-\bar{\rho})=0\ {\rm mod}\ 2\pi i$, which is nothing but Eq.~(2) in the main text.

\subsection{proof of Eq.~(3)}
Next, let us assume the two-fold rotational symmetry $\hat{C}_2$ in addition to $\hat{\tilde{T}}_x$ and $\hat{\tilde{T}}_y$.  Starting from the second line of Eq.~(\ref{sB2}), we can proceed in the same way as the $\pi$-flux case in the main text:
\begin{eqnarray}
2\pi ip L_y C&=&q\oint d\theta_x\int_{-\frac{d}{2}}^{\frac{d}{2}}  d\theta_y\,F^{\bm{\theta}}-i\sum_{m=1}^{q-1}\sum_{j=0}^{m-1}\oint d\theta_x\,\left[n_x^{(\theta_x,-\frac{d}{2}+jd)}-n_x^{(\theta_x,-\frac{d}{2}+(j+1)d)}\right]\notag\\
&=&q\oint d\theta_x\int_{-\frac{d}{2}}^{\frac{d}{2}}  d\theta_y\,F^{\bm{\theta}}-iq\oint d\theta_x\,n_x^{(\theta_x,-\frac{d}{2})}+2\pi iqL_y\bar{\rho}\notag\\
&=&q\oint d\theta_x\int^{\frac{d}{2}}_0d\theta_y\,(F^{\bm{\theta}}+F^{-\bm{\theta}})-iq\oint d\theta_x\,n_x^{(\theta_x,-\frac{d}{2})}+2\pi iqL_y\bar{\rho}\notag\\
&=&2q\oint d\theta_x\int^{\frac{d}{2}}_0 d\theta_y\, F^{\bm{\theta}}-iq\oint d\theta_x (n_x^{(\theta_x,0)}-n_x^{(\theta_x,\frac{d}{2})})-iq\oint d\theta_x\,n_x^{(\theta_x,-\frac{d}{2})}+2\pi iqL_y\bar{\rho}\notag\\
&=&2q\oint d\theta_x\int^{\frac{d}{2}}_0 d\theta_y\, F^{\bm{\theta}}-iq\oint d\theta_x n_x^{(\theta_x,0)}+2\pi iqL_y\bar{\rho}\notag\\
&\overset{(\ast)}{=}&2q\int_{-\pi}^\pi d\theta_x \, (A_x^{(\theta_x,0)}-A_x^{(\theta_x,\frac{d}{2})})-iq\int_{-\pi}^\pi d\theta_x n_x^{(\theta_x,0)}+2\pi iqL_y\bar{\rho}\notag\\
&=&2q\int^{\pi}_{0}d\theta_x[(A_x^{(\theta_x,0)}+A_x^{(-\theta_x,0)})-(A_x^{(\theta_x,\frac{d}{2})}+A_x^{(-\theta_x,\frac{d}{2})})]-2iq\int_{0}^\pi d\theta_x\,n_x^{(\theta_x,0)}+2\pi iqL_y\bar{\rho}\notag\\
&=&2q\int^{\pi}_{0}d\theta_x[(A_x^{(\theta_x,0)}+A_x^{(-\theta_x,0)}-i n_x^{(\theta_x,0)})\notag\\
&&\quad-(A_x^{(\theta_x,\frac{d}{2})}+A_x^{(-\theta_x,-\frac{d}{2})}-i n_x^{(\theta_x,\frac{d}{2})})-(A_x^{(-\theta_x,\frac{d}{2})}-A_x^{(-\theta_x,-\frac{d}{2})}+i n_x^{(-\theta_x,-\frac{d}{2})})]+2\pi iqL_y\bar{\rho}\notag\\
&=&2q\int^{\pi}_{0}d\theta_x\partial_{\theta_x}\ln \frac{w_{C_2}^{(\theta_x,0)}}{w_{T_x}^{(-\theta_x,-\frac{d}{2})}w_{C_2}^{(\theta_x,\frac{d}{2})}}+2\pi iqL_y\bar{\rho}=2q\int^{\pi}_{0}d\theta_x\partial_{\theta_x}\ln \frac{w_{C_2}^{(\theta_x,0)}}{w_{T_xC_2}^{(\theta_x,\frac{d}{2})}}+2\pi iqL_y\bar{\rho}\notag\\
&=&2q\ln\frac{w^{(\pi,0)}_{C_2}w^{(0,\frac{d}{2})}_{T_xC_2}}{w^{(0,0)}_{C_2}w^{(\pi,\frac{d}{2})}_{T_xC_2}}+2\pi iqL_y\bar{\rho}\,.\label{sB3}
\end{eqnarray}
Again the step $(\ast)$ is true only modulo $4\pi iq$. Since $L_y$ is odd and $\frac{p}{q}C-\bar{\rho}$ is an integer, we have $e^{\pi iL_y(\frac{p}{q}C-\bar{\rho})}=e^{\pi i(\frac{p}{q}C-\bar{\rho})}$ and
\begin{equation}
e^{\pi i(\frac{p}{q}C-\bar{\rho})}=\frac{w^{(0,\frac{d}{2})}_{T_xC_2}w^{(\pi,0)}_{C_2}}{w^{(\pi,\frac{d}{2})}_{T_xC_2}w^{(0,0)}_{C_2}}=
w^{(0,0)}_{C_2}w^{(\pi,0)}_{C_2}
w^{(0,\frac{d}{2})}_{T_xC_2}
w^{(\pi,\frac{d}{2})}_{T_xC_2}.
\end{equation}
Writing $\Gamma=(0,0)$, $X=(\pi,0)$, $Y'=(0,\frac{d}{2})$, $M'=(\pi,\frac{d}{2})$ and $T_xC_2=C_2'$, we arrive at Eq.~(3) of the main text.

\clearpage

\section{Appendix D: Proof of Eq.~(1) for $n$-fold rotations ($n=2,3,4,6$)}
\label{app:RotationChern}
In this appendix we prove Eq.~(1) of the main text, relating the many-body Chern number $C$ to the $n$-fold rotation eigenvalues. 
\subsection{Transformation rule of $\theta$ and $A$}
The twisted boundary condition are set by $(\hat{T}_{\vec{a}_1})^L=e^{-i\theta_1\hat{N}}$ and $(\hat{T}_{\vec{a}_2})^L=e^{-i\theta_2\hat{N}}$.
Writing $\vec{R}=m_1\vec{a}_1+m_2\vec{a}_2$ and $\vec{\theta}=\theta_1\vec{b}_1+\theta_2\vec{b}_2$ ($\vec{a}_i\cdot\vec{b}_j=\delta_{i,j}$), we can express the boundary conditions more simply as $(\hat{T}_{\vec{R}})^L=e^{-\vec{\theta}\cdot\vec{R}\hat{N}}$.

Now let us ask how $\vec{\theta}$ transforms under under a rotation operation $\hat{g}$ mapping $\vec{R}\rightarrow p_g\vec{R}$. 
On the one hand, if $\vec{\theta}$ is changed to $\vec{\theta}'$, we have $(\hat{T}_{\vec{R}}')^L=e^{-\vec{\theta}'\cdot\vec{R}\hat{N}}$.  On the other hand, 
\begin{equation}
(\hat{T}_{\vec{R}}')^L=\hat{g}(\hat{T}_{p_g^{-1}\vec{R}})^L\hat{g}^\dagger=\hat{g}(e^{-\vec{\theta}\cdot(p_g^{-1}\vec{R})\hat{N}})^L\hat{g}^{-1}=e^{-(p_g\vec{\theta})\cdot\vec{R}\hat{N}}.
\end{equation}
Therefore, $\vec{\theta}$ transforms as a vector $\vec{\theta}\rightarrow \vec{\theta}'=p_g\vec{\theta}$. In particular, this means that $(\theta_1,\theta_2)$ transforms in the same way as the momenta $(k_1,k_2)$ in $\vec{k}=k_1\vec{b}_1+k_2\vec{b}_2$.  Suppose that the ground state $|\Phi^{\vec{\theta}}\rangle$ is unique and gapped. Then it should satisfy
\begin{equation}
\hat{g}^{\vec{\theta}}|\Phi^{\vec{\theta}}\rangle=w_g^{\vec{\theta}}|\Phi^{p_g\vec{\theta}}\rangle.
\end{equation}
For the product $g=g_2g_1$, we have $\hat{g}_2^{p_{g_1}\vec{\theta}}\hat{g}_1^{\vec{\theta}}=\hat{g}^{\vec{\theta}}$ and $w_{g_2}^{p_{g_1}\vec{\theta}}w_{g_1}^{\vec{\theta}}=w_{g}^{\vec{\theta}}$.

The Berry connection $\vec{A}^{\vec{\theta}}\equiv \langle\Phi^{\vec{\theta}}|\nabla_{\vec{\theta}}|\Phi^{\vec{\theta}}\rangle$ and the Berry cavature $F^{\vec{\theta}}= \nabla_{\vec{\theta}}\times\vec{A}^{\vec{\theta}}$ change to
\begin{eqnarray}
\vec{A}^{p_g\vec{\theta}}&=&p_g\vec{A}^{\vec{\theta}}+p_g\langle\Phi^{\vec{\theta}}|[(\hat{g}^{\vec{\theta}})^\dagger\nabla_{\vec{\theta}}\hat{g}^{\vec{\theta}}]|\Phi^{\vec{\theta}}\rangle-p_g\nabla_{\vec{\theta}}\ln w_g^{\vec{\theta}},\label{app:transA}\\
F^{p_g\vec{\theta}}&=&F^{\vec{\theta}}+ \nabla_{\vec{\theta}}\times\langle\Phi^{\vec{\theta}}|[(\hat{g}^{\vec{\theta}})^\dagger\nabla_{\vec{\theta}}\hat{g}^{\vec{\theta}}]|\Phi^{\vec{\theta}}\rangle.
\end{eqnarray}
As we will see shortly, $(\hat{g}^{\vec{\theta}})^\dagger\nabla_{\vec{\theta}}\hat{g}^{\vec{\theta}}$ is a local charge operator and we will write
\begin{equation}
\langle\hat{n}\rangle_{\vec{x}}^{\vec{\theta}}\equiv\langle\Phi^{\vec{\theta}}|\hat{n}_{\vec{x}}|\Phi^{\vec{\theta}}\rangle.
\end{equation}
The rotation symmetry implies that
\begin{equation}
\langle\hat{n}\rangle_{\vec{x}}^{\vec{\theta}}=\langle\Phi^{\vec{\theta}}|\hat{n}_{\vec{x}}|\Phi^{\vec{\theta}}\rangle=\langle\Phi^{\vec{\theta}}|(\hat{g}^{\vec{\theta}})^\dagger\hat{g}^{\vec{\theta}}\hat{n}_{\vec{x}}(\hat{g}^{\vec{\theta}})^\dagger\hat{g}^{\vec{\theta}}|\Phi^{\vec{\theta}}\rangle=\langle\hat{n}\rangle_{p_g\vec{x}}^{p_g\vec{\theta}}.
\end{equation}

\begin{figure}[b]
\begin{center}
\includegraphics[width=0.6\columnwidth]{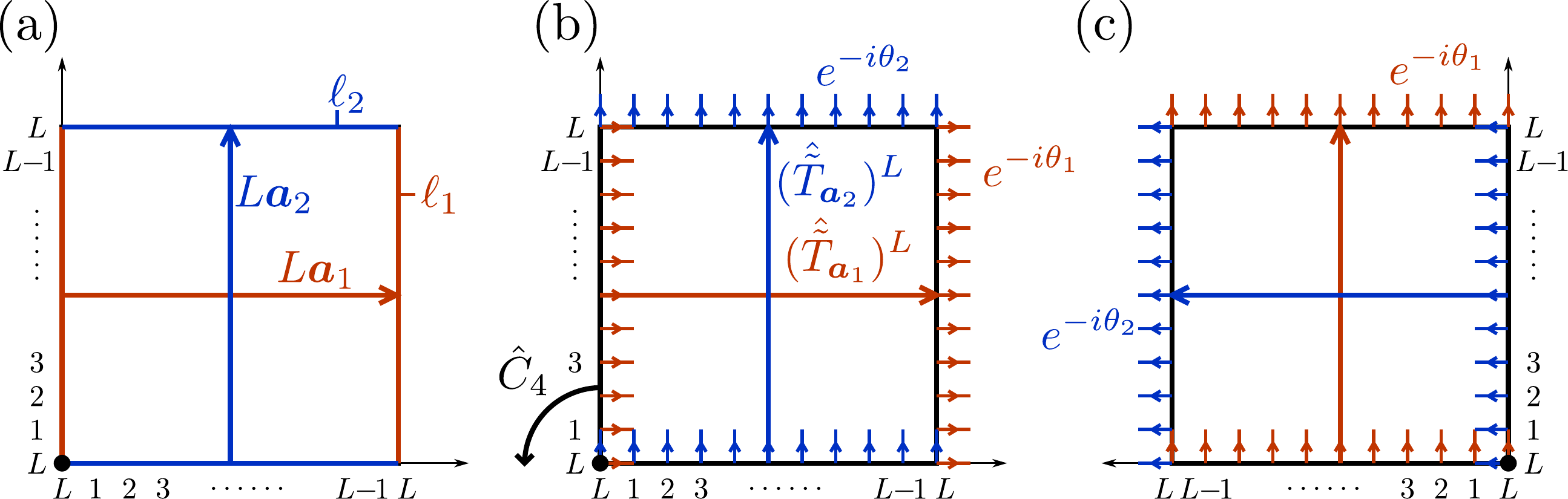}
\caption{(a, b) Illustration of the twisted boundary condition for $C_4$ symmetric models. Twisted boundary conditions introduce phase factor $e^{-\theta_i}$ along the line $\ell_i$ ($i=1,2$). (c) The $C_4$ rotation not only changes $(\theta_1,\theta_2)$ to $(-\theta_2,\theta_1)$ but also shifts the position of the twisted bond with $e^{-i\theta_2}$ by one unit. We need the phase rotation $e^{-i\theta_2\sum_{\vec{x}\in\ell_2}^L\hat{n}_{\vec{x}}}$ in Eq.~\eqref{app:c4op} to fix the position.
\label{app:figc4}}
\end{center}
\end{figure}

\subsection{$C_2$, $C_4$ rotation}

We write $\vec{a}_1=(1,0)$, $\vec{a}_2=(0,1)$, $\vec{b}_1=(1,0)$, $\vec{b}_2=(0,1)$, $\vec{x}=x_1\vec{a}_1+x_2\vec{a}_2$, and $\vec{\theta}=\theta_1\vec{b}_1+\theta_2\vec{b}_2$.  Since 
\begin{eqnarray}
C_4\vec{x}=x_1\vec{a}_2+x_2(-\vec{a}_1)=-x_2\vec{a}_1+x_1\vec{a}_2,\\
C_4\vec{\theta}=\theta_1\vec{b}_2+\theta_2(-\vec{b}_1)=-\theta_2\vec{b}_1+\theta_1\vec{b}_2,
\end{eqnarray}
we see $(x_1,x_2)\mapsto(-x_2,x_1)$ and $(\theta_1,\theta_2)\mapsto(-\theta_2,\theta_1)$ under $C_4$. 

As illustrated in Fig.~\ref{app:figc4}, twisted rotations are given by
\begin{eqnarray}
\hat{C}_4^{(\theta_1,\theta_2)}&=&\hat{P}_4e^{-i\theta_2\sum_{\vec{x}\in\ell_2}^L\hat{n}_{\vec{x}}}=\hat{P}_4e^{-i\theta_2\sum_{x_1=1}^L\hat{n}_{(x_1,L)}},\label{app:c4op}\\
\hat{C}_2^{(\theta_1,\theta_2)}&=&\hat{C}_4^{(-\theta_2,\theta_1)}\hat{C}_4^{(\theta_1,\theta_2)}=\hat{P}_2e^{-i\theta_1\sum_{\vec{x}\in\ell_1}^L\hat{n}_{\vec{x}}-i\theta_2\sum_{\vec{x}\in\ell_2}^L\hat{n}_{\vec{x}}}=\hat{P}_2e^{-i\theta_1\sum_{x_2=1}^L\hat{n}_{(L,x_2)}-i\theta_2\sum_{x_1=1}^L\hat{n}_{(x_1,L)}}.
\end{eqnarray}
Assuming that the ground state $|\Phi^{(\theta_1,\theta_2)}\rangle$ is unique and gapped for all values of $\theta_{1,2}$, we have
\begin{eqnarray}
\hat{C}_4^{(\theta_1,\theta_2)}|\Phi^{(\theta_1,\theta_2)}\rangle&=&w_{C_4}^{(\theta_1,\theta_2)}|\Phi^{(-\theta_2,\theta_1)}\rangle,\\
A_1^{(-\theta_1,-\theta_2)}&=&-A_1^{(\theta_1,\theta_2)}+i\sum_{x_2=1}^L\langle\hat{n}\rangle_{(L,x_2)}^{(\theta_1,\theta_2)}+\partial_{\theta_1}\ln w_{C_2}^{(\theta_1,\theta_2)},\\
A_1^{(-\theta_2,\theta_1)}&=&-A_2^{(\theta_1,\theta_2)}+i\sum_{x_1=1}^L\langle\hat{n}\rangle_{(x_1,L)}^{(\theta_1,\theta_2)}+\partial_{\theta_2}\ln w_{C_4}^{(\theta_1,\theta_2)},\\
A_2^{(-\theta_2,\theta_1)}&=&A_1^{(\theta_1,\theta_2)}-\partial_{\theta_1}\ln w_{C_4}^{(\theta_1,\theta_2)},
\end{eqnarray}

The high-symmetric values of $\vec{\theta}=\theta_1\vec{b}_1+\theta_2\vec{b}_2$ are 
\begin{equation}
\textstyle\Gamma=(0,0),\quad X=\pi\vec{b}_1, \quad Y=\pi\vec{b}_2,\quad M=\pi(\vec{b}_1+\vec{b}_2).
\end{equation}

\begin{figure}[b]
\begin{center}
\includegraphics[width=0.5\columnwidth]{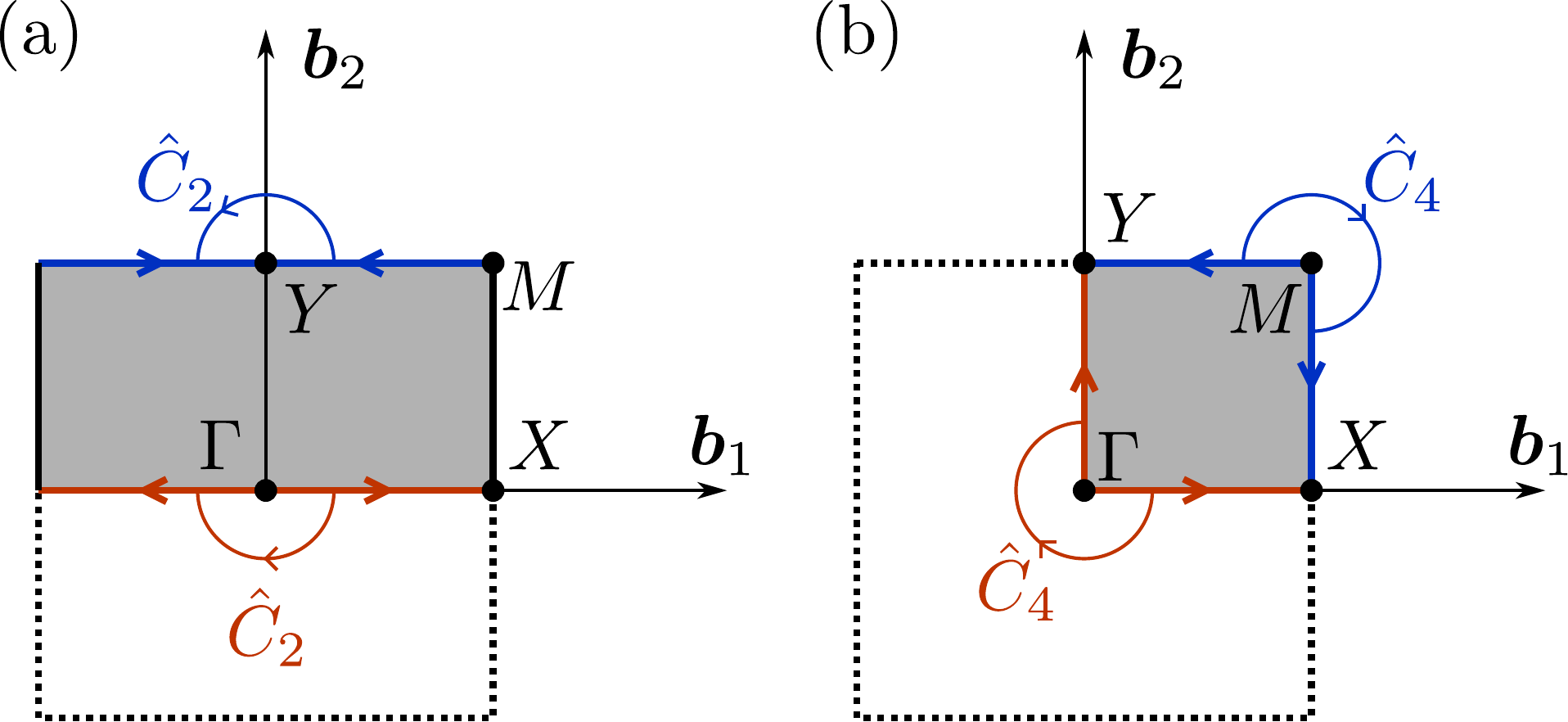}
\caption{Irreducible part of $\theta_1\vec{b}_1+\theta_2\vec{b}_2$ for (a) $C_2$ and (b) $C_4$ symmetric case.  The arrows in the same color on the edge of the irreducible part are related by the transformation rule in Eq.~\eqref{app:transA}.
\label{app:figbzc4}}
\end{center}
\end{figure}

\subsubsection{$C_2$ rotation}
For $C_2$ rotation, we have [see Fig.~\ref{app:figbzc4} (a)]
\begin{eqnarray}
&&2\pi iC=\oint d\theta_1\oint d\theta_2 F^{(\theta_1,\theta_2)}=\oint d\theta_1\int_{0}^{\pi} d\theta_2 [F^{(\theta_1,\theta_2)}+F^{(-\theta_1,-\theta_2)}]\notag\\
&=&2\oint d\theta_1\int_{0}^{\pi} d\theta_2 F^{(\theta_1,\theta_2)}-i\oint d\theta_1\sum_{x_2=1}^L\left[\langle\hat{n}\rangle_{(L,x_2)}^{(\theta_1,0)}-\langle\hat{n}\rangle_{(L,x_2)}^{(\theta_1,\pi)}\right]\notag\\
&\overset{\text{(mod $4\pi i$)}}{=}&2\oint d\theta_1 [A_1^{(\theta_1,0)}-A_1^{(\theta_1,\pi)}]-i\int_0^\pi d\theta_1\sum_{x_2=1}^L\left[\langle\hat{n}\rangle_{(L,x_2)}^{(\theta_1,0)}+\langle\hat{n}\rangle_{(L,x_2)}^{(-\theta_1,0)}-\langle\hat{n}\rangle_{(L,x_2)}^{(\theta_1,\pi)}-\langle\hat{n}\rangle_{(L,x_2)}^{(-\theta_1,\pi)}\right]\notag\\
&=&2\int_{0}^{\pi} d\theta_1 [A_1^{(\theta_1,0)}+A_1^{(-\theta_1,0)}-A_1^{(\theta_1,\pi)}-A_1^{(-\theta_1,\pi)}]-2i\int_{0}^\pi d\theta_1\sum_{x_2=1}^L\left[\langle\hat{n}\rangle_{(L,x_2)}^{(\theta_1,0)}-\langle\hat{n}\rangle_{(L,x_2)}^{(\theta_1,\pi)}\right]\notag\\
&=&2\int_{0}^{\pi} d\theta_1 [\partial_{\theta_1}\ln w_{C_2}^{(\theta_1,0)}-\partial_{\theta_1}\ln w_{C_2}^{(\theta_1,\pi)}]=2\ln \frac{w_{C_2}^{(\pi,0)}w_{C_2}^{(0,\pi)}}{w_{C_2}^{(0,0)}w_{C_2}^{(\pi,\pi)}}=2\ln \frac{w_{C_2}^{X}w_{C_2}^{Y}}{w_{C_2}^{\Gamma}w_{C_2}^{M}}.\label{appC2int}
\end{eqnarray}
Therefore,
\begin{equation}
e^{\frac{2\pi i}{2}C}=(-1)^C=\frac{w_{C_2}^{X}w_{C_2}^{Y}}{w_{C_2}^{\Gamma}w_{C_2}^{M}}\quad\Leftrightarrow\quad e^{-\frac{2\pi i}{2}C}=w_{C_2}^{\Gamma}w_{C_2}^{X}w_{C_2}^{Y}w_{C_2}^{M}.\label{app:c2}
\end{equation}

\subsubsection{$C_4$ rotation}
For $C_4$ rotation, we start from the second line of Eq.~\eqref{appC2int}. As illustrated in Fig.~\ref{app:figbzc4} (b), we can further halve the integration range.
\begin{eqnarray}
&&2\pi iC=2\oint d\theta_1\int_{0}^{\pi} d\theta_2 F^{(\theta_1,\theta_2)}-i\oint d\theta_1\sum_{x_2=1}^L\left[\langle\hat{n}\rangle_{(L,x_2)}^{(\theta_1,0)}-\langle\hat{n}\rangle_{(L,x_2)}^{(\theta_1,\pi)}\right]\notag\\
&=&2\int_0^\pi d\theta_1\int_{0}^{\pi} d\theta_2 [F^{(\theta_1,\theta_2)}+F^{(-\theta_2,\theta_1)}]-2i\int_{0}^\pi d\theta\sum_{x=1}^L\left[\langle\hat{n}\rangle_{(L,x)}^{(\theta,0)}-\langle\hat{n}\rangle_{(L,x)}^{(\theta,\pi)}\right]\notag\\
&=&4\int_0^\pi d\theta_1\int_{0}^{\pi} d\theta_2 F^{(\theta_1,\theta_2)}
-2i\int_0^\pi d\theta \sum_{x=1}^L\left[\langle\hat{n}\rangle_{(x,L)}^{(\pi,\theta)}-\langle\hat{n}\rangle_{(x,L)}^{(0,\theta)}\right]-2i\int_{0}^\pi d\theta\sum_{x=1}^L\left[\langle\hat{n}\rangle_{(L,x)}^{(\theta,0)}-\langle\hat{n}\rangle_{(L,x)}^{(\theta,\pi)}\right]\notag\\
&\overset{\text{(mod $8\pi i$)}}{=}&4\int_0^\pi d\theta [A_1^{(\theta,0)}-A_1^{(\theta,\pi)}+A_2^{(\pi,\theta)}-A_2^{(0,\theta)}]=4\int_{0}^{\pi} d\theta [\partial_{\theta}\ln w_{C_4}^{(\theta,0)}-\partial_{\theta}\ln w_{C_4}^{(\theta,\pi)}]=4\frac{w_{C_4}^{Y}w_{C_4}^{X}}{w_{C_4}^{\Gamma}w_{C_4}^{M}}.
\end{eqnarray}
Therefore,
\begin{equation}
e^{\frac{2\pi i}{4}C}=i^C=\frac{w_{C_2}^{X}}{w_{C_4}^{\Gamma}w_{C_4}^{M}}\quad\Leftrightarrow\quad e^{-\frac{2\pi i}{4}C}=(-1)^{2SN}w_{C_4}^{\Gamma}w_{C_2}^{X}w_{C_4}^{M}.
\end{equation}
In the last step, we used $(w_{C_2}^{X})^2=w_{(C_2)^2}^{X}=(-1)^{2SN}$.

\subsection{$C_6$, $C_3$ rotation}

\begin{figure}
\begin{center}
\includegraphics[width=0.8\columnwidth]{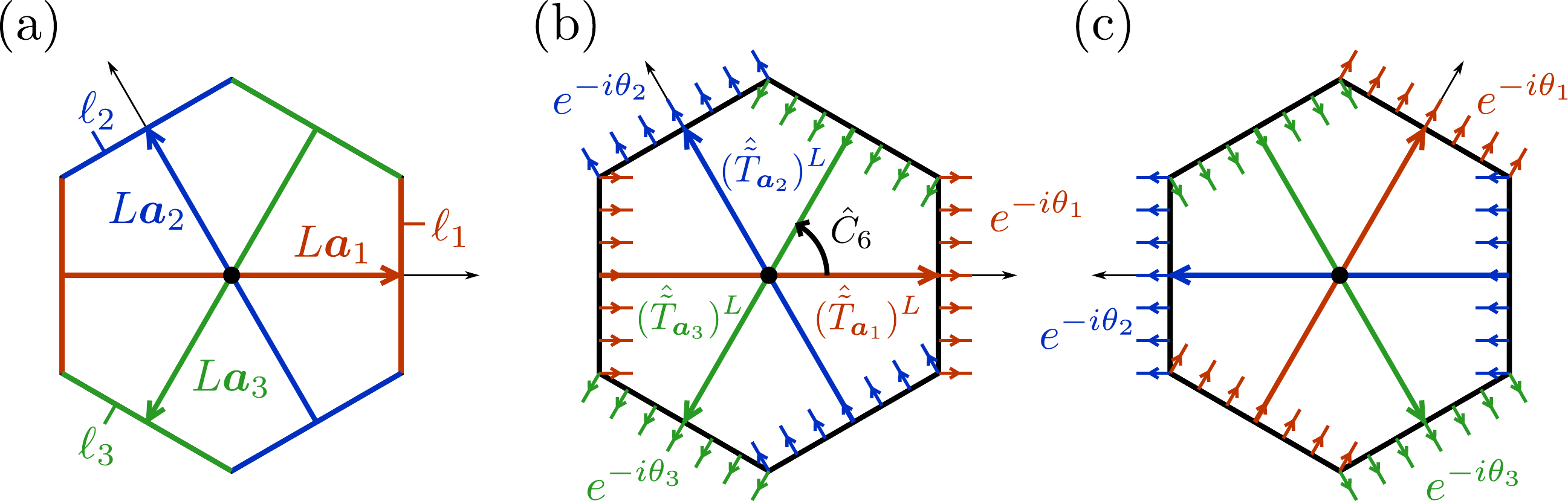}
\caption{(a, b) Illustration of the twisted boundary condition for $C_6$ symmetric models. $\theta_3\equiv-(\theta_1+\theta_2)$ as explained in the text. Twisted boundary conditions introduce phase factor $e^{-\theta_i}$ along the line $\ell_i$ ($i=1,2,3$). (c) The $C_6$ rotation not only changes $(\theta_1,\theta_2,\theta_3)$ to $(-\theta_2,-\theta_3,-\theta_1)$ but also shifts the position of the twisted bond. We need the phase rotation $ \hat{u}_6^{(\theta_1,\theta_2)}= e^{-i\theta_1\sum_{\vec{x}\in \ell_1}\hat{n}_{\vec{x}}-i\theta_2\sum_{\vec{x}\in \ell_2}\hat{n}_{\vec{x}}-i\theta_3\sum_{\vec{x}\in \ell_3}\hat{n}_{\vec{x}}}$ in Eq.~\eqref{app:c6op} to fix the position.
\label{app:figc6}}
\end{center}
\end{figure}

We write $\vec{a}_1=(1,0)$, $\vec{a}_2=(-\frac{1}{2},\frac{\sqrt{3}}{2})$, $\vec{b}_1=(1,\frac{1}{\sqrt{3}})$, $\vec{b}_2=(0,\frac{2}{\sqrt{3}})$, $\vec{x}=x_1\vec{a}_1+x_2\vec{a}_2$, and $\vec{\theta}=\theta_1\vec{b}_1+\theta_2\vec{b}_2$. 
\begin{eqnarray}
C_6\vec{x}=x_1(\vec{a}_1+\vec{a}_2)+x_2(-\vec{a}_1)=(x_1-x_2)\vec{a}_1+x_1\vec{a}_2,\\
C_6\vec{\theta}=\theta_1\vec{b}_2+\theta_2(\vec{b}_2-\vec{b}_1)=-\theta_2\vec{b}_1+(\theta_1+\theta_2)\vec{b}_2,
\end{eqnarray}
we see $(x_1,x_2)\mapsto(x_1-x_2,x_1)$ and $(\theta_1,\theta_2)\mapsto(-\theta_2,\theta_1+\theta_2)$ under $C_6$.  

Another way of deriving the transformation rule of $\theta$ is the following. It is convenient to introduce $\vec{a}_3\equiv -(\vec{a}_1+\vec{a}_2)$ as illustrated in Fig.~\ref{app:figc6}. The conditions $(\hat{T}_{\vec{a}_1})^L=e^{-i\theta_1\hat{N}}$ and $(\hat{T}_{\vec{a}_2})^L=e^{-i\theta_2\hat{N}}$ imply that $(\hat{T}_{\vec{a}_3})^L=e^{-i\theta_3\hat{N}}$ with $\theta_3\equiv -(\theta_1+\theta_2)$.
The transformation rule of $(\theta_1,\theta_2,\theta_3)$ can be determined by
\begin{eqnarray}
e^{i\theta_1'\hat{N}}&=&(\hat{T}_{\vec{a}_1}^{-L})'=\hat{C}_6(\hat{T}_{\vec{a}_2})^L\hat{C}_6^{-1}=e^{-i\theta_2\hat{N}},\\
e^{i\theta_2'\hat{N}}&=&(\hat{T}_{\vec{a}_2}^{-L})'=\hat{C}_6(\hat{T}_{\vec{a}_3})^L\hat{C}_6^{-1}=e^{-i\theta_3\hat{N}},\\
e^{i\theta_3'\hat{N}}&=&(\hat{T}_{\vec{a}_3}^{-L})'=\hat{C}_6(\hat{T}_{\vec{a}_1})^L\hat{C}_6^{-1}=e^{-i\theta_1\hat{N}},
\end{eqnarray}
suggesting that $(\theta_1,\theta_2,\theta_3)\mapsto(-\theta_2,-\theta_3,-\theta_1)$ under $C_6$. Therefore, $(\theta_1,\theta_2)\mapsto(-\theta_2,\theta_1+\theta_2)$ under $C_6$ and $(\theta_1,\theta_2)\mapsto(-\theta_1-\theta_2,\theta_1)$ under $C_3=(C_6)^2$. 

As shown in Fig.~\ref{app:figc6}, twisted rotations are given by
\begin{eqnarray}
\hat{C}_6^{(\theta_1,\theta_2)}&=&\hat{P}_6\hat{u}_6^{(\theta_1,\theta_2)},\quad \hat{u}_6^{(\theta_1,\theta_2)}\equiv e^{-i\theta_1\sum_{\vec{x}\in \ell_1}\hat{n}_{\vec{x}}-i\theta_2\sum_{\vec{x}\in \ell_2}\hat{n}_{\vec{x}}-i\theta_3\sum_{\vec{x}\in \ell_3}\hat{n}_{\vec{x}}},\label{app:c6op}\\
\hat{C}_3^{(\theta_1,\theta_2)}&=&\hat{C}_6^{(-\theta_2,\theta_1+\theta_2)}\hat{C}_6^{(\theta_1,\theta_2)}=\hat{P}_3,
\end{eqnarray}
The high-symmetric values of $\vec{\theta}=\theta_1\vec{b}_1+\theta_2\vec{b}_2$ are 
\begin{equation}
\textstyle\Gamma=(0,0),\quad K=\frac{2\pi}{3}\vec{b}_1+\frac{2\pi}{3}\vec{b}_2, \quad K'=\frac{4\pi}{3}\vec{b}_1-\frac{2\pi}{3}\vec{b}_2,\quad M_1=\pi \vec{b}_1,\quad M_2=\pi\vec{b}_2,\quad M_3=\pi(\vec{b}_1+\vec{b}_2).
\end{equation}

\subsubsection{$C_3$ rotation}
Since $\hat{C}_3^{(\theta_1,\theta_2)}$ does not depend on $\theta_{1,2}$, 
\begin{eqnarray}
2\pi iC=\int_{\frac{1}{3}\text{BZ}}d^2\theta [F^{(\theta_1,\theta_2)}+F^{C_3(\theta_1,\theta_2)}+F^{C_3^2(\theta_1,\theta_2)}]=3\int_{\frac{1}{3}\text{BZ}}d^2\theta F^{(\theta_1,\theta_2)}\overset{\text{(mod $6\pi i$)}}{=}3\ln \frac{w_{(C_3)^2}^{K'}}{w_{C_3}^{\Gamma}w_{C_3}^{K}},
\end{eqnarray}
where $\frac{1}{3}\text{BZ}$ is the shaded region in Fig.~\ref{app:figbzc6} (a).  Therefore,
\begin{equation}
e^{\frac{2\pi i}{3}C}=\omega^C=\frac{w_{(C_3)^2}^{K'}}{w_{C_3}^{\Gamma}w_{C_3}^{K}}\quad\Leftrightarrow\quad e^{-\frac{2\pi i}{3}C}=(-1)^{2SN}w_{C_3}^{\Gamma}w_{C_3}^{K}w_{C_3}^{K'}.\label{app:c3}
\end{equation}
In the last step, we used $w_{C_3}^{K'}w_{(C_3)^2}^{K'}=w_{(C_3)^3}^{K'}=(-1)^{2SN}$.

\subsubsection{$C_6$ rotation}
The formula for $C_6$ rotation can be readily derived by combining Eqs.~\eqref{app:c2} and \eqref{app:c3}. The high-symmetry points $X=\pi\vec{b}_1$, $Y=\pi\vec{b}_2$, $M=\pi(\vec{b}_1+\vec{b}_2)$ in Eq.~\eqref{app:c2} should be interpreted as $M_1=\pi \vec{b}_1$, $M_2=\pi\vec{b}_2$, $M_3=\pi(\vec{b}_1+\vec{b}_2)$, respectively.  Since $\frac{1}{6}=\frac{1}{2}-\frac{1}{3}$, we have
\begin{eqnarray}
e^{-\frac{2\pi i}{6}C}&=&e^{-(\frac{2\pi i}{2}C-\frac{2\pi i}{3}C)}=\frac{w_{C_2}^{\Gamma}w_{C_2}^{M_1}w_{C_2}^{M_2}w_{C_2}^{M_3}}{(-1)^{2SN}w_{C_3}^{\Gamma}w_{C_3}^{K}w_{C_3}^{K'}}\notag\\
&=&(-1)^{2SN}(w_{C_2}^{\Gamma}w_{C_3^{-1}}^{\Gamma})(w_{C_2}^{K'}w_{C_3^{-1}}^{K'})(w_{C_2}^{K}w_{C_3^{-1}}^{K})\frac{(w_{C_2}^{M})^3}{w_{C_2}^{K'}w_{C_2}^{K}}\notag\\
&=&(-1)^{2SN}w_{C_6}^{\Gamma}w_{C_6}^{K'}w_{C_6}^{K}\frac{w_{(C_2)^3}^{M}}{w_{(C_2)^2}^{K}}=(-1)^{2SN}w_{C_6}^{\Gamma}w_{C_3}^{K}w_{C_2}^{M}.
\end{eqnarray}
In the derivation, we used $w_{C_2}^{M_1}=w_{C_2}^{M_2}=w_{C_2}^{M_3}$ $(\equiv w_{C_2}^{M})$ as they are all symmetry related and rotations about $z$-axis commutes.

\begin{figure}
\begin{center}
\includegraphics[width=0.45\columnwidth]{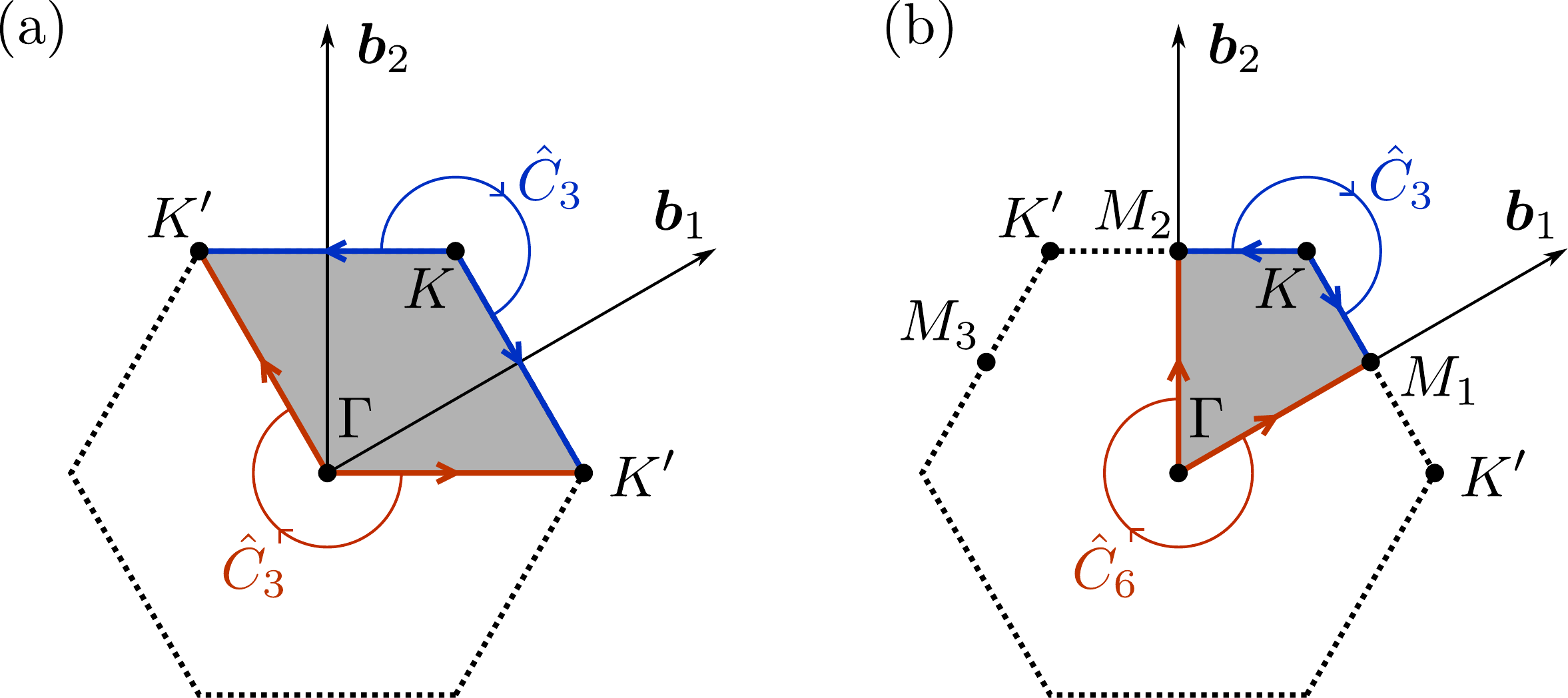}
\caption{Irreducible part of $\theta_1\vec{b}_1+\theta_2\vec{b}_2$ for (a) $C_3$ and (b) $C_6$ symmetric case. 
\label{app:figbzc6}}
\end{center}
\end{figure}

Alternatively, one can derive the same result through an explicit yet tedious calculations. To simplify equations let us introduce
\begin{equation}
-i\langle n\rangle^{(\theta_1(t),\theta_2(t))}=\langle\Phi^{(\theta_1(t),\theta_2(t))}|(\hat{u}_6^{(\theta_1(t),\theta_2(t))})^\dagger\left(\tfrac{d}{dt}\hat{u}_6^{(\theta_1(t),\theta_2(t))}\right)|\Phi^{(\theta_1(t),\theta_2(t))}\rangle.
\end{equation}
Here, $(\theta_1(t),\theta_2(t))$ stands for a line connecting two high-symmetry points parametrized by $t\in[0,1]$.  Using the $C_6$ rotation symmetry we can halve the integration range to $\frac{1}{6}\text{BZ}$ shown in Fig.~\ref{app:figbzc6} (b).
\begin{eqnarray}
2\pi iC&=&6\int_{\frac{1}{6}\text{BZ}}d^2\theta F^{(\theta_1,\theta_2)}\notag\\
&&-3i\int_{0}^1 dt\,\left[\langle n\rangle^{\frac{t}{2}\vec{b}_1}-\langle n\rangle^{\frac{t}{3}\vec{b}_1+\frac{t}{3}\vec{b}_2}+\langle n\rangle^{\frac{3-t}{6}\vec{b}_1+\frac{t}{3}\vec{b}_2}\right]\notag-3i\int_{0}^1 dt\,\left[-\langle n\rangle^{\frac{t}{2}\vec{b}_2}+\langle n\rangle^{\frac{t}{3}\vec{b}_1+\frac{t}{3}\vec{b}_2}-\langle n\rangle^{\frac{t}{3}\vec{b}_1+\frac{3-t}{6}\vec{b}_2}\right]\notag\\
&=&6\int_0^1dt\left[
\tfrac{1}{2}\vec{b}_1\cdot\vec{A}^{\frac{t}{2}\vec{b}_1}
+(-\tfrac{1}{6}\vec{b}_1+\tfrac{1}{3}\vec{b}_2)\cdot\vec{A}^{\frac{3-t}{6}\vec{b}_1+\frac{t}{3}\vec{b}_2}
-(\tfrac{1}{3}\vec{b}_1-\tfrac{1}{6}\vec{b}_2)\cdot\vec{A}^{\frac{t}{3}\vec{b}_1+\frac{3-t}{6}\vec{b}_2}
-\tfrac{1}{2}\vec{b}_2\cdot\vec{A}^{\frac{t}{2}\vec{b}_2}
\right]\notag\\
&&-3i\int_{0}^1 dt\,\left[\langle n\rangle^{\frac{t}{2}\vec{b}_1}-\langle n\rangle^{\frac{t}{2}\vec{b}_2}+\langle n\rangle^{\frac{3-t}{6}\vec{b}_1+\frac{t}{3}\vec{b}_2}-\langle n\rangle^{\frac{t}{3}\vec{b}_1+\frac{3-t}{6}\vec{b}_2}\right]\notag\\
&=&6\int_0^1dt\left[
i\langle n\rangle^{\frac{t}{2}\vec{b}_1}+\partial_t \ln w_{C_6}^{\frac{t}{2}\vec{b}_1}-\partial_t \ln w_{C_3}^{\frac{t}{3}\vec{b}_1+\frac{3-t}{6}\vec{b}_2}
\right]-3i\int_{0}^1 dt\,\left[\langle n\rangle^{\frac{t}{2}\vec{b}_1}-\langle n\rangle^{\frac{t}{2}\vec{b}_2}+\langle n\rangle^{\frac{3-t}{6}\vec{b}_1+\frac{t}{3}\vec{b}_2}-\langle n\rangle^{\frac{t}{3}\vec{b}_1+\frac{3-t}{6}\vec{b}_2}\right]\notag\\
&=&3i\int_0^1dt\left[
\langle n\rangle^{\frac{t}{2}\vec{b}_1}+\langle n\rangle^{\frac{t}{2}\vec{b}_2}
\right]+6\ln\frac{w_{C_3}^{M_2}w_{C_6}^{M_1}}{w_{C_6}^{\Gamma}w_{C_3}^{K}}=6\ln\frac{w_{C_2}^{M_1}}{w_{C_6}^{\Gamma}w_{C_3}^{K}}
\end{eqnarray}
where we used
\begin{eqnarray}
\langle n\rangle^{\frac{3-t}{6}\vec{b}_1+\frac{t}{3}\vec{b}_2}-\langle n\rangle^{\frac{t}{3}\vec{b}_1+\frac{3-t}{6}\vec{b}_2}=0,\quad \langle n\rangle^{\frac{t}{2}\vec{b}_1}+\langle n\rangle^{\frac{t}{2}\vec{b}_2}=0,
\end{eqnarray}
each of which follows from $C_6$ rotation symmetry. Therefore,
\begin{equation}
e^{\frac{2\pi i}{6}C}=\frac{w_{C_2}^{M}}{w_{C_6}^{\Gamma}w_{C_3}^{K}}\quad\Leftrightarrow\quad e^{-\frac{2\pi i}{6}C}=(-1)^{2SN}w_{C_6}^{\Gamma}w_{C_3}^{K}w_{C_2}^{M}.
\end{equation}

\clearpage

\section{Appendix E: $C_4$-rotation eigenvalues and many-body Chern number under magnetic field}
\label{app:C4pi}
In this appendix we demonstrate that our approach equally works for higher rotations, i.e. $C_n$ ($n=3,4,6$), using a few examples for $n=4$.
\subsection{$\frac{\pi}{2}$ flux}
\begin{figure}[b]
\begin{center}
\includegraphics[width=0.8\columnwidth]{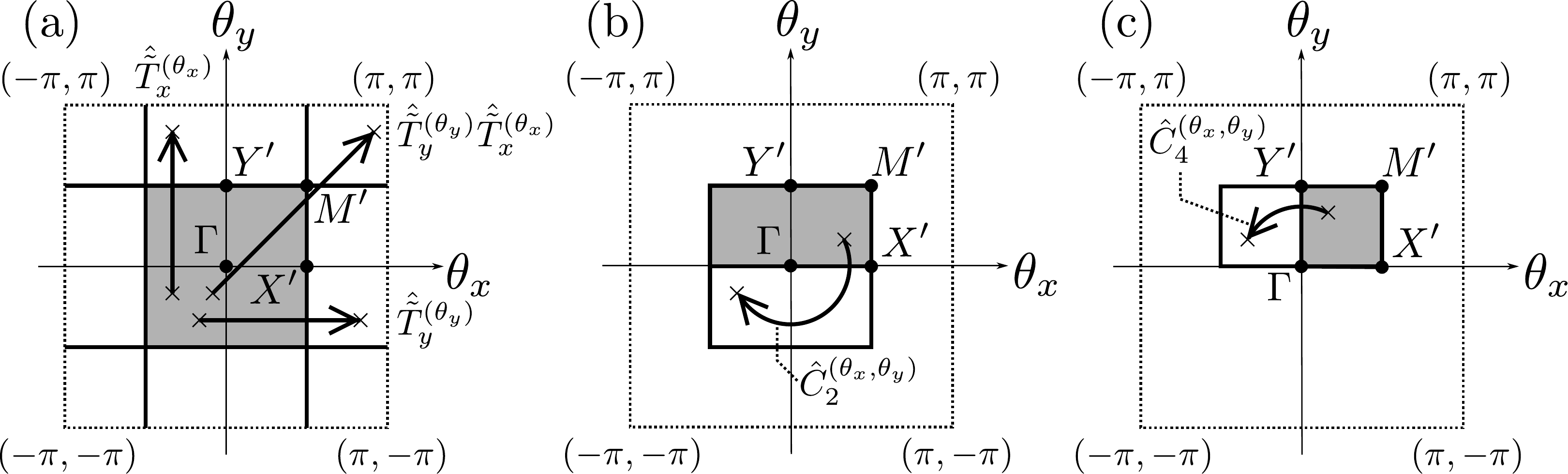}
\caption{The idea of the derivation in Eq.~\eqref{app:pi2c42}. Translation symmetries $\tilde{T}_x$, $\tilde{T}_y$ and the rotation symmetry $C_4$ makes the irreducible part 16 times smaller.
\label{app:figpi2c4}}
\end{center}
\end{figure}

Let us start with the simplest case of $\phi=\frac{\pi}{2}$. The four fold rotation symmetry sets $L_x=L_y=L$.  In order to include an integer multiple of $2\pi$ flux in total, $L$ has to be even but $L/2$ can be odd.
With this choice, $\hat{\tilde{T}}_x^{\theta_x}$ ($\hat{\tilde{T}}_y^{\theta_y}$) shifts $\theta_y$ ($\theta_x$) by $\pi$.  Therefore,  $\hat{\tilde{T}}_x^{\theta_x}$ and $\hat{\tilde{T}}_y^{\theta_y}$ individually halves the irreducible part of the integration range [Fig.~\ref{app:figpi2c4} (a)]. On the top of it, the four fold rotation further cut it down, leaving only $\frac{1}{2}\times\frac{1}{2}\times\frac{1}{4}=\frac{1}{16}$ size of the ``Brillouin zone" [Fig.~\ref{app:figpi2c4} (b,c)].

Indeed,
\begin{eqnarray}
&&2\pi i C=\int_{-\pi}^{\pi}d\theta_x\int_{-\pi}^{\pi}d\theta_yF^{\vec{\theta}}=\int_{\frac{-\pi}{2}}^{\frac{\pi}{2}}d\theta_x\int_{\frac{-\pi}{2}}^{\frac{\pi}{2}}d\theta_y[F^{\vec{\theta}}+F^{(\theta_x+\pi,\theta_y)}+F^{(\theta_x,\theta_y+\pi)}+F^{(\theta_x+\pi,\theta_y+\pi)}]\notag\\
&=&4\int_{\frac{-\pi}{2}}^{\frac{\pi}{2}}d\theta_x\int_{\frac{-\pi}{2}}^{\frac{\pi}{2}}d\theta_yF^{\vec{\theta}}-2i\sum_{x=1}^{L}\int_{\frac{-\pi}{2}}^{\frac{\pi}{2}}d\theta\left[\langle\hat{n}\rangle_{(L,x)}^{(\theta,\frac{-\pi}{2})}-\langle\hat{n}\rangle_{(L,x)}^{(\theta,\frac{\pi}{2})}+\langle\hat{n}\rangle_{(x,L)}^{(\frac{\pi}{2},\theta)}-\langle\hat{n}\rangle_{(x,L)}^{(\frac{-\pi}{2},\theta)}\right]\notag\\
&=&4\int_{\frac{-\pi}{2}}^{\frac{\pi}{2}}d\theta_x\int_{0}^{\frac{\pi}{2}}d\theta_y[F^{\vec{\theta}}+F^{-\vec{\theta}}]\notag\\
&=&8\int_{\frac{-\pi}{2}}^{\frac{\pi}{2}}d\theta_x\int_{0}^{\frac{\pi}{2}}d\theta_yF^{\vec{\theta}}-4i\sum_{x=1}^{L}\int_{0}^{\frac{\pi}{2}}d\theta[
\langle\hat{n}\rangle_{(L,x)}^{(\theta,0)}+\langle\hat{n}\rangle_{(L,x)}^{(-\theta,0)}-\langle\hat{n}\rangle_{(L,x)}^{(\theta,\frac{\pi}{2})}-\langle\hat{n}\rangle_{(L,x)}^{(-\theta,\frac{\pi}{2})}
+\langle\hat{n}\rangle_{(x,L)}^{(\frac{\pi}{2},\theta)}-\langle\hat{n}\rangle_{(x,L)}^{(\frac{-\pi}{2},\theta)}]\notag\\
&=&8\int_{0}^{\frac{\pi}{2}}d\theta_x\int_{0}^{\frac{\pi}{2}}d\theta_y[F^{\vec{\theta}}+F^{(-\theta_y,\theta_x)}]-8i\sum_{x=1}^{L}\int_{0}^{\frac{\pi}{2}}d\theta[\langle\hat{n}\rangle_{(L,x)}^{(\theta,0)}-\langle\hat{n}\rangle_{(L,x)}^{(\theta,\frac{\pi}{2})}]\notag\\
&=&16\int_{0}^{\frac{\pi}{2}}d\theta_x\int_{0}^{\frac{\pi}{2}}d\theta_yF^{\vec{\theta}}
-8i\sum_{x=1}^{L}\int_{0}^{\frac{\pi}{2}}d\theta [\langle\hat{n}\rangle_{(x,L)}^{(\frac{\pi}{2},\theta)}-\langle\hat{n}\rangle_{(x,L)}^{(0,\theta)}]
-8i\sum_{x=1}^{L}\int_{0}^{\frac{\pi}{2}}d\theta[\langle\hat{n}\rangle_{(L,x)}^{(\theta,0)}-\langle\hat{n}\rangle_{(L,x)}^{(\theta,\frac{\pi}{2})}]\notag\\
&=&16\int_{0}^{\frac{\pi}{2}}d\theta_x\int_{0}^{\frac{\pi}{2}}d\theta_yF^{\vec{\theta}}
-8i\sum_{x=1}^{L}\int_{0}^{\frac{\pi}{2}}d\theta [\langle\hat{n}\rangle_{(x,L)}^{(\frac{\pi}{2},\theta)}-\langle\hat{n}\rangle_{(L,x)}^{(\theta,\frac{\pi}{2})}]\notag\\
&\overset{\text{(mod $32\pi i$)}}{=}&16\int_{0}^{\frac{\pi}{2}}d\theta (A_x^{(\theta,0)}-A_y^{(0,\theta)}-A_x^{(\theta,\frac{\pi}{2})}+A_y^{(\frac{\pi}{2},\theta)})
-8i\sum_{x=1}^{L}\int_{0}^{\frac{\pi}{2}}d\theta [\langle\hat{n}\rangle_{(x,L)}^{(\frac{\pi}{2},\theta)}-\langle\hat{n}\rangle_{(x,L)}^{(-\frac{\pi}{2},\theta)}]\notag\\
&=&16\ln\frac{w_{C_4}^{(\frac{\pi}{2},0)}w_{T_yC_4}^{(0,\frac{\pi}{2})}}{w_{C_4}^{(0,0)}w_{T_yC_4}^{(\frac{\pi}{2},\frac{\pi}{2})}}-8i\sum_{x=1}^{L}\int_{0}^{\frac{\pi}{2}}d\theta [\langle\hat{n}\rangle_{(x,L)}^{(\frac{\pi}{2},\theta)}+\langle\hat{n}\rangle_{(x,L)}^{(-\frac{\pi}{2},\theta)}]=16\ln\frac{w_{C_4}^{(\frac{\pi}{2},0)}w_{T_yC_4}^{(0,\frac{\pi}{2})}}{w_{C_4}^{(0,0)}w_{T_yC_4}^{(\frac{\pi}{2},\frac{\pi}{2})}}-8i\pi L\bar{\rho}.\label{app:pi2c42}
\end{eqnarray}
Therefore,
\begin{equation}
e^{\frac{2\pi i}{4}(\frac{1}{4}C+L\bar{\rho})}=\frac{w_{T_yC_2}^{(\frac{\pi}{2},0)}}{w_{C_4}^{(0,0)}w_{T_yC_4}^{(\frac{\pi}{2},\frac{\pi}{2})}}.
\end{equation}
In the derivation, we used
\begin{eqnarray}
A_x^{(\theta_x,\theta_y+\pi)}&=&A_x^{\vec{\theta}}-i\sum_{y=1}^{L}\langle\hat{n}\rangle_{(L,y)}^{\vec{\theta}}-\partial_{\theta_x}\ln\omega_{T_x}^{\vec{\theta}},\\
A_y^{(\theta_x,\theta_y+\pi)}&=&A_y^{\vec{\theta}}-\partial_{\theta_y}\ln\omega_{T_x}^{\vec{\theta}},\\
A_x^{(\theta_x+\pi,\theta_y)}&=&A_x^{\vec{\theta}}-\partial_{\theta_x}\ln\omega_{T_y}^{\vec{\theta}},\\
A_y^{(\theta_x+\pi,\theta_y)}&=&A_y^{\vec{\theta}}-i\sum_{x=1}^{L}\langle\hat{n}\rangle_{(x,L)}^{\vec{\theta}}-\partial_{\theta_y}\ln\omega_{T_y}^{\vec{\theta}},
\end{eqnarray}
and
\begin{eqnarray}
A_y^{(0,\theta)}&=&A_x^{(\theta,0)}-\partial_{\theta}\ln w_{C_4}^{(\theta,0)}.\\
A_y^{(\frac{\pi}{2},\theta)}&=&A_x^{(\theta,\frac{\pi}{2})}-i\sum_{x=1}^{L}\langle\hat{n}\rangle_{(x,L)}^{(-\frac{\pi}{2},\theta)}-\partial_{\theta}\ln w_{T_yC_4}^{(\theta,\frac{\pi}{2})}.
\end{eqnarray}

Clearly, one can perform the same calculation for $\phi=\frac{2\pi}{n^2}$ flux ($n\geq1$) by choosing $L_x=L_y=L$ to be an integer multiple of $n$.
\subsection{$\pi$ flux}
\begin{figure}
\begin{center}
\includegraphics[width=0.8\columnwidth]{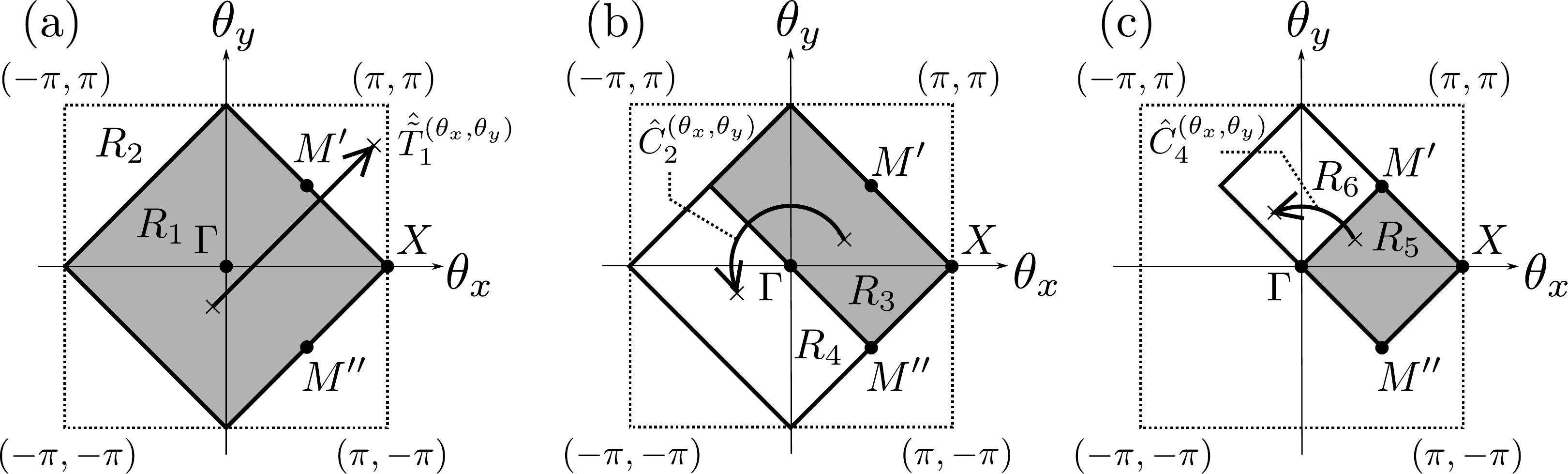}
\caption{The sketch of the caculation in Eqs.~\eqref{app:pic41} and \eqref{app:pic42}. Translation symmetry $\tilde{T}_1=\tilde{T}_x\tilde{T}_y$ and the rotation symmetry $C_4$ makes the irreducible part 8 times smaller.
\label{app:figpic4}}
\end{center}
\end{figure}

Here let us discuss the next simplest example of $\phi=\pi$ in order to show that one can also deal with fluxes other than $\phi=\frac{2\pi}{n^2}$ in a way consistent with the four-fold rotation symmetry.  
To this end, we use the same operators as in the previous section for $\phi=\frac{\pi}{2}$ but assume that the Hamiltonian only symmetric under
\begin{equation}
(\hat{\tilde{T}}_x^{\theta_x})^{2},\quad (\hat{\tilde{T}}_y^{\theta_y})^{2},\quad \hat{\tilde{T}}_x^{(\theta_x+\pi,\theta_y)}\hat{\tilde{T}}_y^{\vec{\theta}},
\end{equation}
and their products.  $(\hat{\tilde{T}}_x^{\theta_x})^{2}$ and $(\hat{\tilde{T}}_y^{\theta_y})^{2}$ do not change $(\theta_x,\theta_y)$, while $\hat{\tilde{T}}_x^{(\theta_x+\pi,\theta_y)}\hat{\tilde{T}}_y^{\vec{\theta}}$ shifts $(\theta_x,\theta_y)$ by $(\pi,\pi)$. Unlike the previous calculation, we do not individually assume $\hat{\tilde{T}}_x^{\theta_x}$ and $\hat{\tilde{T}}_y^{\theta_y}$ and the irreducible part is only 8 times smaller than the original integration range (Fig.~\ref{app:figpic4}).   We have
\begin{eqnarray}
2\pi iC&=&\oint d\theta_x\oint d\theta_y F^{\vec{\theta}}=\int_{R_1}d^2\theta [F^{\vec{\theta}}+F^{(\theta_x+\pi,\theta_y+\pi)}]\notag\\
&=&2\int_{R_1}d^2\theta\,F^{\vec{\theta}}-i\sum_{x=1}^{L}\int_{\partial R_1}d\vec{\theta}\cdot\left(
      \langle\hat{n}\rangle_{(L,x)}^{\vec{\theta}},
      \langle\hat{n}\rangle_{(x,L)}^{\vec{\theta}} 
  \right)\notag\\
&=&2\int_{R_3}d^2\theta[F^{\vec{\theta}}+F^{-\vec{\theta}}]\notag\\
&=&4\int_{R_3}d^2\theta\,F^{\vec{\theta}}-2i\sum_{x=1}^{L}\int_{\partial R_3}d\vec{\theta}\cdot\left(
      \langle\hat{n}\rangle_{(L,x)}^{\vec{\theta}},
      \langle\hat{n}\rangle_{(x,L)}^{\vec{\theta}} 
  \right)\notag\\
&=&4\int_{R_5}d^2\theta[F^{\vec{\theta}}+F^{(-\theta_y,\theta_x)}]-4i\sum^{L}_{x=1}\int^{\frac{\pi}{2}}_0d\theta\big[\langle\hat{n}\rangle_{(x,L)}^{(\theta,\theta)} -\langle\hat{n}\rangle_{(x,L)}^{(\theta,-\theta)}+\langle\hat{n}\rangle_{(x,L)}^{(\theta,-\theta+\pi)}-\langle\hat{n}\rangle_{(x,L)}^{(\theta,\theta-\pi)} \big]\notag\\
&=&8\int_{R_5}d^2\theta\,F^{\vec{\theta}}-4i\sum_{x=1}^{L}\int_{\partial R_5}d\vec{\theta}\cdot\left(
      0,
      \langle\hat{n}\rangle_{(x,L)}^{\vec{\theta}} 
      \right)
-4i\sum^{L}_{x=1}\int^{\frac{\pi}{2}}_0d\theta\big[\langle\hat{n}\rangle_{(x,L)}^{(\theta,\theta)} -\langle\hat{n}\rangle_{(x,L)}^{(\theta,-\theta)}+\langle\hat{n}\rangle_{(x,L)}^{(\theta,-\theta+\pi)}-\langle\hat{n}\rangle_{(x,L)}^{(\theta,\theta-\pi)} \big]\notag\\
&=&8\int_{R_5}d^2\theta F^{\vec{\theta}}-4i\sum_{x=1}^{L}\int^{\frac{\pi}{2}}_0d\theta\big[-2\langle\hat{n}\rangle_{(x,L)}^{(\theta,-\theta)}+\langle\hat{n}\rangle_{(x,L)}^{(\theta,-\theta+\pi)}+\langle\hat{n}\rangle_{(x,L)}^{(-\theta+\pi,-\theta)}+\langle\hat{n}\rangle_{(x,L)}^{(-\theta+\pi,\theta)}-\langle\hat{n}\rangle_{(x,L)}^{(\theta,\theta-\pi)}\big.]\label{app:pic41}
\end{eqnarray}
See Fig.~\ref{app:figpic4} for the defintions of the regions $R_n$ ($n=1,2,\ldots 6$).  Again using the Stoke's theorem,
\begin{eqnarray}
&&8\int_{R_5}d^2\theta F^{\vec{\theta}}\overset{\text{(mod $16\pi i$)}}{=}8\int_{\partial R_5}d\vec{\theta}\cdot\vec{A}^{\vec{\theta}}\notag\\
&=&8\int^{\frac{\pi}{2}}_{0}d\theta[A_x^{(\theta,-\theta)}-A_y^{(\theta,\theta)}-A_y^{(\theta,-\theta)}-A_x^{(\theta,\theta)}+A_x^{(-\theta+\pi,-\theta)}+A_y^{(-\theta+\pi,\theta)}+A_y^{(-\theta+\pi,-\theta)}-A_x^{(-\theta+\pi,\theta)}]\notag\\
&=&8\ln\frac{w_{C_4^{-1}}^{(0,0)}w_{T_xT_yC_4^{-1}}^{(\pi,0)}}{w_{C_4^{-1}}^{(\frac{\pi}{2},\frac{\pi}{2})}w_{T_xT_yC_4^{-1}}^{(\frac{\pi}{2},\frac{-\pi}{2})}}-8i\sum_{x=1}^{L}\int^{\frac{\pi}{2}}_{0}d\theta[\langle\hat{n}\rangle_{(x,L)}^{(\theta,-\theta)}-\langle\hat{n}\rangle_{(x,L)}^{(\theta-\pi,\theta)}].\label{app:pic42}
\end{eqnarray}
The second term in the last line of Eq.~\eqref{app:pic41} cancels against the second term also in the last line Eq.~\eqref{app:pic42}. Therefore,
\begin{eqnarray}
e^{\frac{2\pi i}{8}C}=\frac{w_{C_4^{-1}}^{(0,0)}w_{T_xT_yC_4^{-1}}^{(\pi,0)}}{w_{C_4^{-1}}^{(\frac{\pi}{2},\frac{\pi}{2})}w_{T_xT_yC_4^{-1}}^{(\frac{\pi}{2},\frac{-\pi}{2})}}=\frac{w_{C_4^{-1}}^{(0,0)}w_{T_xT_yC_4^{-1}}^{(\pi,0)}}{w_{T_xT_yC_2}^{(\frac{\pi}{2},\frac{\pi}{2})}}=\frac{w_{C_4^{-1}}^{\Gamma}w_{T_xT_yC_4^{-1}}^{X}}{w_{T_xT_yC_2}^{M'}}.
\end{eqnarray}
There are several other equivalent ways to express the same result. For example,
\begin{eqnarray}
e^{i\frac{2\pi}{4}(\frac{1}{2}C+L\bar{\rho})}=\frac{w_{T_xT_yC_2}^{(\frac{\pi}{2},\frac{-\pi}{2})}}{w_{C_4}^{(0,0)}w_{T_xT_yC_4}^{(\pi,0)}}=\frac{w_{T_xT_yC_2}^{M''}}{w_{C_4}^{\Gamma}w_{T_xT_yC_4}^{X}}.
\end{eqnarray}

\clearpage

\section{Appendix F: Fractional Quantum Hall States}
\label{app:FQHS}

In the main text, we assumed the uniqueness of the ground state on the twisted torus by setting $\nu\equiv q\bar{\rho}$ to be an integer. In order to include fractional quantum Hall states, here let us set $\nu = p'/q'$ ($p'$ and $q'$ are co-prime) and assume that there exists $D=q'$ degenerate ground states below the excitation gap.  

To proceed, let us further assume $L_y$ to be co-prime with $q$ and $q'$. Then $L_x$ must be an integer multiple of $qq'$ so that the total flux $\phi L_xL_y=L_xL_yp/q$ is an integer multiple of $2\pi$ and that the total number of particles $N=\bar{\rho}L_xL_y=L_xL_yp'/qq'$ is an integer. Given this choice, we will first show that $q'$ ground states can be obtained by starting from one ground state $\ket{\Phi^{\vec{\theta}}}$ and changing $\theta_x$ by $2\pi n$ ($n=1,2,\ldots,q'-1$).

Since $(\hat{\tilde{T}}_x^{\theta_x})^q$ commutes with the Hamiltonian $\hat{H}^{\vec{\theta}}$, $\ket{\Phi^{\vec{\theta}}}$  can be chosen simultaneously an eigenstate of $(\hat{\tilde{T}}_x^{\theta_x})^q$ and let us denote the eigenvalue by $w_{T_x^q}^{\vec{\theta}}$.  As we impose the twisted boundary condition $(\hat{\tilde{T}}_x^{\theta_x})^{L_x}=e^{-i\theta_xN}$, the eigenvalue  must satisfy
\begin{equation}
(w_{T_x^q}^{\vec{\theta}})^{L_x/q}=e^{-i\theta_xN}.
\end{equation}
Recalling that $N=\bar{\rho}L_xL_y=(L_x/q)(p'L_y/q')$, we can write
\begin{equation}
w_{T_x^q}^{\vec{\theta}}=e^{-i\theta_xp'L_y/q'}e^{2\pi i n_0 q/L_x}
\end{equation}
with an integer $1\leq n_0\leq L_x/q$.  

Now let us smoothly change $\theta_x$.  As $\theta_x$ is increased by $2\pi$, $w_{T_x^q}^{\vec{\theta}}$ acquires a
phase $e^{-2\pi i p'L_y/q'}\neq1$, implying that the state $\ket{\Phi^{(\theta_x+2\pi,\theta_y)}}$ is different from $\ket{\Phi^{\vec{\theta}}}$. Further assuming that the gap above the $D$ ground states does not vanish for any value of $\theta_x$ and $\theta_y$, we can see that $\ket{\Phi^{(\theta_x+2\pi,\theta_y)}}$ is one of the $D$ degenerate ground states.   One can repeat this argument and prove that the following $D$ states are all distinct ground states of the Hamiltonian $\hat{H}^{\vec{\theta}}$:
\begin{eqnarray}
\ket{\Phi^{\vec{\theta}}}, \ket{\Phi^{(\theta_x+2\pi,\theta_y)}},\cdots, \ket{\Phi^{(\theta_x+2\pi(D-1),\theta_y)}}.
\end{eqnarray}

In this case, the Hall conductivity is given by $\tilde{\sigma}_{xy}=\frac{e^2}{2\pi\hbar}\tilde{C}$, where
\begin{eqnarray}
\tilde{C}=\frac{1}{2\pi iD}\sum_{n=0}^{D-1}\int^{2\pi}_{0} d\theta_x\oint d\theta_y F^{(\theta_x+2\pi n,\theta_y)}=\frac{1}{2\pi iD}\int^{2\pi D}_{0}d\theta_x\oint d\theta_y F^{\vec{\theta}}.
\end{eqnarray}
By performing the same calculation as in the main text, we obtain
\begin{eqnarray}
e^{2\pi i\left(\frac{p}{q}\tilde{C}-\bar{\rho}\right)D}=1.
\end{eqnarray}

\clearpage

\end{document}